\documentclass[12pt,a4paper]{iopart}
\pdfoutput=1
\usepackage{iopams,setstack,mathrsfs,amsthm,cite,enumerate,float,graphicx}
\usepackage[colorlinks,linkcolor=blue,citecolor=blue,urlcolor=blue]{hyperref}
\newcommand{\al}{\alpha}

\newcommand{\de}{\delta}

\newcommand{\vep}{\varepsilon}
\newcommand{\ga}{\gamma}

\newcommand{\la}{\lambda}

\newcommand{\si}{\sigma}
\renewcommand{\th}{\theta}
\newcommand{\vp}{\varphi}

\newcommand{\bev}{\mathbf e}
\newcommand{\bh}{\mathbf{h}}
\newcommand{\biv}{\mathbf{i}} 
\newcommand{\bj}{\mathbf{j}} 

\newcommand{\bn}{\mathbf{n}}

\newcommand{\bt}{\mathbf{t}} 

\newcommand{\bx}{\mathbf{x}}
\newcommand{\by}{\mathbf{y}}

\newcommand{\bS}{\mathbf S}

\newcommand{\NN}{{\mathbb N}}
\newcommand{\RR}{{\mathbb R}}
\newcommand{\ZZ}{{\mathbb Z}}
\newcommand{\cE}{{\mathcal E}}

\newcommand{\cH}{{\mathcal H}}

\newcommand{\cN}{{\mathcal N}}

\newcommand{\cS}{{\mathcal S}}

\newcommand{\pd}{\partial}

\newcommand{\ket}[1]{|#1\rangle}
\newcommand{\bra}[1]{\langle#1|} 

\newcommand{\mss}{\kern 1pt}
\renewcommand{\leq}{\leqslant}

\renewcommand{\le}{\leqslant}
\renewcommand{\ge}{\geqslant}
\newcommand{\tends}[1]{\bbuildrel{\hbox to 2em{\rightarrowfill}}_{#1}^{}}
\newcommand{\inner}[2]{\langle #1|#2\rangle}
\let\iu\rmi
\let\diff\rmd
\newcommand{\su}{\mathrm{su}}

\newcommand{\en}{\enspace}

\newcommand{\pdf}[2]{\frac{\partial #1}{\partial #2}}

\newcommand{\Int}[1]{\,\mathop{\!#1}\limits^{\lower1ex\hbox{$\scriptstyle\circ$}}{}}

\theoremstyle{remark}
\newtheorem{rem}{Remark}

\let\tfrac\case
\let\eqref\eref

\newcommand{\binom}[2]{{#1\choose #2}}
\newcommand{\mathclap}[1]{\hbox to0pt{\hss$\scriptstyle #1$\hss}}
\newcommand{\hps}[1]{\hphantom{\hbox{$\scriptstyle #1$}}}
\eqnobysec
\begin{document}

\title[Generalized LMG models: ground state entanglement and quantum entropies]{Generalized
  isotropic Lipkin--Meshkov--Glick models: ground state entanglement and quantum entropies}

\author{Jos\'e A.~Carrasco$^1$, Federico Finkel$^1$, Artemio Gonz\'alez-L\'opez$^1$,
  Miguel A.~Rodr\'iguez$^1$ and Piergiulio Tempesta$^{1,2}$}

\address{$^1$Departamento de F\'{\i}sica Te\'{o}rica II, Universidad Complutense de Madrid, 28040
  Madrid, Spain}
\address{$^2$Instituto de Ciencias
  Matem\'aticas, c/ Nicol\'as Cabrera n\textordmasculine{} 13--15, 28049 Madrid, Spain}

 \eads{\mailto{joseacar@ucm.es}, \mailto{ffinkel@ucm.es},
  \mailto{artemio@ucm.es}, \mailto{rodrigue@ucm.es} and \mailto{p.tempesta@fis.ucm.es,
    piergiulio.tempesta@icmat.es}}

\date{January 11, 2016}

\begin{abstract}
  We introduce a new class of generalized isotropic Lipkin--Meshkov--Glick models with $\su(m+1)$
  spin and long-range non-constant interactions, whose non-degenerate ground state is a Dicke
  state of~$\su(m+1)$ type. We evaluate in closed form the reduced density matrix of a block of
  $L$ spins when the whole system is in its ground state, and study the corresponding von Neumann
  and R\'enyi entanglement entropies in the thermodynamic limit. We show that both of these
  entropies scale as $a\log L$ when~$L$ tends to infinity, where the coefficient~$a$ is equal to
  $(m-k)/2$ in the ground state phase with $k$ vanishing $\su(m+1)$ magnon densities. In
  particular, our results show that none of these generalized Lipkin--Meshkov--Glick models are
  critical, since when~$L\to\infty$ their R\'enyi entropy $R_q$ becomes independent of the
  parameter~$q$. We have also computed the Tsallis entanglement entropy of the ground state of
  these generalized~$\su(m+1)$ Lipkin--Meshkov--Glick models, finding that it can be made
  extensive by an appropriate choice of its parameter only when $m-k\ge3$. Finally, in
  the~$\su(3)$ case we construct in detail the phase diagram of the ground state in parameter
  space, showing that it is determined in a simple way by the weights of the fundamental
  representation of~$\su(3)$. This is also true in the $\su(m+1)$ case; for instance, we prove
  that the region for which all the magnon densities are non-vanishing is an~$(m+1)$-simplex
  in~$\RR^m$ whose vertices are the weights of the fundamental representation of~$\su(m+1)$.
\end{abstract}
{\it Keywords\/}: entanglement in extended quantum systems (theory), solvable
lattice models, conformal field theory (theory)

\maketitle

\section{Introduction}

A crucial difference between classical and quantum systems is the fact
that in the latter ones the entropy of a subsystem can be positive even when the whole system is
in its (pure) ground state at zero temperature. Indeed, classically a positive value of the
entropy is simply a reflection of the lack of knowledge of the precise microstate of the system
when it is in a certain macrostate. On the other hand, an essential property of quantum systems is
the fact that the exact knowledge of the state of the whole system does not give complete
information about the state of a subsystem. This is a consequence of the entanglement among
different parts of the system, which is perhaps the most paradigmatic quantum
phenomenon~\cite{HHHH09,NC10}.

One of the most widespread quantitative measures of the degree of entanglement of a quantum system
at zero temperature is the bipartite entropy of its ground state~$\ket\psi$. More precisely, if we
divide the system into two subsystems with~$L$ and~$N-L$ particles the corresponding bipartite
entropy is the quantum entropy of the reduced density matrix~$\rho_L=\tr_{N-L}\rho$,
where~$\rho\equiv\ket\psi\bra\psi$ and the trace is taken over the degrees of freedom of the
latter subsystem. This definition is actually symmetric between both subsystems, since the entropy
of~$\rho_L$ necessarily coincides with that of~$\rho_{N-L}$. In practice, the von
Neumann~\cite{HHHH09} and R\'enyi~\cite{Re61,Re70} entropies, respectively defined by
\begin{equation}
  S=-\tr(\rho_L\log\rho_L),\qquad
  R_q=\frac{\log\tr(\rho_L^q)}{1-q}
  \label{VN-Renyi}
\end{equation}
(where $q$ is a positive real parameter) have been extensively used as quantitative measures of
the bipartite entanglement. The exact evaluation of these entropies, or even the determination of
their large $L$ limit, is only possible for a handful of mostly one-dimensional models. These
models include the XX and XY nearest-neighbors Heisenberg spin chains, for which the asymptotic
behavior of both the von Neumann and R\'enyi entropies was established in the last
decade~\cite{VLRK03,JK04,IJK05} with the help of the Fisher--Hartwig conjecture. As is well known,
both of these models are critical in a certain region of their parameter space~\cite{Sa11}. The
large~$L$ behavior of the entropy is essentially different in the critical and non-critical
regions. More precisely, the entropy scales as $\log L$ in the critical region, while it saturates
to a constant in the non-critical one. This is a manifestation of the so-called area
law~\cite{ECP10}, according to which the (von Neumann) bipartite entropy of a critical
one-dimensional quantum system with short-range interactions behaves as $\log L$, while for
non-critical systems it tends to a constant. This is consistent with the fact that the bipartite
entropy of two-dimensional conformal field theories (which describe one-dimensional quantum
critical systems in the thermodynamic limit) scales as $(c+\bar c)(1+q^{-1})(\log L)/12$,
where~$c$ and~$\bar c$ denote respectively the holomorphic and anti-holomorphic central charges
and~$q=1$ for the von Neumann entropy~\cite{HLW94,CC04JSTAT,CC05}.

On the other hand, it is generally believed that the area law need not hold for systems exhibiting
long-range interactions, since the stronger correlations in these systems are expected to increase
the entropy. One of the few one-dimensional long-range quantum systems that has been exactly
solved is the Lipkin--Meshkov--Glick (LMG) model~\cite{LMG65,MGL65,GLM65}, originally introduced
by the latter authors to describe a system of $N$ fermions in two levels each of which is $N$
times degenerate. This model is equivalent to a system of $N$ spin $1/2$ particles with constant
XY-type long-range interactions in a transverse magnetic field. In the isotropic (XX) case the
model is exactly solvable, and its bipartite von Neumann entropy has also been computed in closed
form~\cite{PS05,LORV05}. The von Neumann entropy vanishes in the gapped (and not entangled) phase,
while it grows as $(\log L)/2$ in the gapless one. Thus the LMG model behaves in this respect as a
one-dimensional quantum system with \emph{short}-range interactions. The reason of this behavior
is the competition between the range of the interactions, which tends to increase the entropy, and
the high degree of symmetry of this model, which tends to lower it. Indeed, in this case the
ground state is a symmetric Dicke state (i.e., it has maximum total spin $S=N/2$ and a
well-defined value of the total spin component~$S^z$) for all values of its parameters, which is
easily seen to imply that the entanglement entropy cannot exceed~$\log(L+1)$.

A class of one-dimensional models with long-range interactions which has been extensively studied
is that of spin chains of Haldane--Shastry (HS) type. The original~HS chain~\cite{Ha88,Sh88} is a
lattice model of $N$ spin~$1/2$ particles uniformly arranged on a circle, with pairwise
interactions inversely proportional to the square of the chord distance. This model was soon
afterwards generalized to particles carrying $\su(m+1)$ spin, as well as to rational and
hyperbolic interactions~\cite{Fr93,Po93,Po94,FI94}. The HS chain, which is closely related to the
one-dimensional Hubbard model with long-range hopping~\cite{GR92}, is exactly solvable via the
asymptotic Bethe ansatz, and its partition function can be evaluated in closed form~\cite{FG05}.
Among other remarkable properties, the HS chain possesses exact Yangian symmetry for a finite
number of sites~\cite{HHTBP92}, and provides one of the simplest realizations of anyons in one
dimension via Haldane's fractional statistics~\cite{Ha91b,HHTBP92,BGHP93,Gr09}.

In this paper we shall introduce a large family of $\su(m+1)$ spin chains which, like the HS-type
chains, feature variable long-range interactions, and whose ground state entanglement properties
are similar to those of the isotropic LMG model. More precisely, the models we shall construct
will admit as non-degenerate ground state a generalized Dicke state of~$\su(m+1)$ type, i.e., a
state totally symmetric under permutations and with a well defined number of particles in each of
the~$\su(m+1)$ internal one-particle states ($\su(m+1)$ magnons). This will be achieved by
replacing the quadratic term in the total spin operator $S^z$ present in the~LMG Hamiltonian by a
sum of similar terms in each of the generators of the~$\su(m+1)$ Cartan subalgebra. The resulting
models can thus be considered a natural generalization of the original (isotropic) LMG model, to
which they actually reduce when $m=1$ and all the two-body interactions are constant.

An explicit expression for the reduced density matrix~$\rho_L$ of any system whose ground state is
an $\su(m+1)$-type Dicke state with arbitrary $m$ first appeared in Ref.~\cite{PSS05}. The
eigenvalues of~$\rho_L$ turn out to define a hypergeometric distribution in $m$ variables, thus
generalizing the result of Refs.~\cite{PS05,LORV05} for the spin $1/2$ ($m=1$) isotropic LMG
model. We shall show that in the thermodynamic limit~$N\gg1$ with $L/N\to\al$ (finite) the
eigenvalues of the reduced density matrix can be well approximated by a Gaussian distribution,
whose parameters we evaluate in closed form for arbitrary~$m$ and~$\al$. With the help of this
approximation, we obtain explicit asymptotic expressions for the von Neumann and R\'enyi entropies
in the thermodynamic limit. In particular, our expression for the von Neumann entropy coincides
with that of Ref.~\cite{PSS05}, derived by extrapolation from the~$\al=0$ case. Remarkably, in the
region of parameter space for which all the ground state magnon densities are non-vanishing both
of these entropies scale as $(m\log L)/2$ as~$L$ tends to infinity. Thus the behavior of the von
Neumann entropy is that of a critical model with~$c=\bar c=3m/2$, described by a conformal field
theory (CFT) with~$m$ fermions and $m$ bosons. This may not seem surprising at first sight, taking
into account that many critical one-dimensional spin chains, including the Heisenberg (XXX) and
the~$\su(m+1)$~HS chains, are effectively described by theories of this type~(see, e.g.,
\cite{HHTBP92,Af85,Sc94}). Here, however, the situation is more subtle. Indeed, the R\'enyi
entropy~$R_q$, although still proportional to~$\log L$ for large~$L$, becomes independent of the
parameter~$q$ for~$L\to\infty$, and as a consequence the family of generalized isotropic LMG
models cannot contain any critical instances.

It is well known that a crucial requirement of classical thermodynamics is the extensivity of the
(Maxwell--Boltzmann) entropy of a given system, i.e., that $S\propto L^{d}$, where $L$ is a
characteristic length of the system and $d$ is the number of space dimensions. In a quantum
context this requirement, at least for the von Neumann entropy, is violated in many cases, as for
instance in black hole thermodynamics~\cite{Be73,Be74,Ha74,Ha76}. In fact, the area law mentioned
above evidences a non-extensive behavior of von Neumann's entropy in strongly correlated quantum
systems. On the other hand, it is very natural to enquire whether this feature is shared by all
quantum entropies available in the literature. Interestingly enough, this is not the case. For
instance, as already noted in Ref.~\cite{CT08}, the quantum Tsallis entropy~\cite{Ts88,Ts09} can
be extensive when von Neumann's is not in several one- and two-dimensional strongly correlated
systems, which include the Heisenberg~XY model. We have found a similar behavior for the Tsallis
entanglement entropy of the ground state of generalized $\su(m+1)$ LMG models with $m-k\ge3$
(where $k$ is the number of vanishing magnon densities), while for $m-k=1,2$ the Tsallis entropy
is not extensive for any value of its parameter.

As mentioned above, the ground state of the isotropic LMG model has two quantum phases (entangled
and non-entangled), respectively determined by the values of the (suitably normalized) magnetic
field strength being less or greater than $1$ in absolute value. For the generalized LMG models
constructed in this paper, the situation is more subtle. More precisely, we shall show that in
this case the ground state can be in exactly $m+1$ quantum phases, each of which is characterized
by the vanishing of a certain number of magnon densities. Moreover, in the phase with $k$
vanishing magnon densities both the von Neumann and R\'enyi entropies scale as~$\frac12(m-k)\log L$,
implying again that none of these phases can contain any critical models. We have performed a
detailed analysis of the ground state phases in the $\su(3)$ ($m=2$) case, completely identifying
the corresponding regions in parameter space. Remarkably, these regions are entirely determined in
a geometric way by the weights of the fundamental representation of~$\su(3)$ associated to the
choice of the Cartan generators. A similar result holds in the general ($\su(m+1)$) case; for
instance, we show that the region for which all the magnon densities are non-vanishing is an
$(m+1)$-simplex in~$\RR^m$ whose vertices are the weights of the fundamental representation of
$\su(m+1)$.

We end this section with a few words on the paper's organization. In Section~\ref{sec.LMG} we
review the main properties of the isotropic LMG model and present the construction of
its~$\su(m+1)$ generalization with non-constant interactions. In Section~\ref{sec.entropies} we
evaluate in closed form the ground state entanglement entropies of von Neumann and R\'enyi, and
study their main properties. As a byproduct of our analysis, we obtain a similar expression for
the Tsallis entropy and discuss its extensivity. In Section~\ref{sec.gspd}, a detailed description
of the entanglement properties of the ground state as a function of the $\su(m+1)$ magnetic field
strength is presented. Special attention is paid to the case $m=2$, for which we obtain a complete
phase diagram describing the regions in parameter space with different magnon content.
Section~\ref{sec.conc} is devoted to the presentation of our conclusions and the discussion of
some open problems suggested by the previous results. The paper ends with three technical
appendices in which we present a detailed derivation of the exact formula for the eigenvalues of
the reduced density matrix, compute the first and second moments of a multivariate hypergeometric
distribution, and show how a univariate hypergeometric distribution can be approximated by a
normal one in a suitable limit.

\section{Generalized Lipkin--Meshkov--Glick models}\label{sec.LMG}

The original Lipkin--Meshkov--Glick model describes a system of $N$ mutually interacting spin 1/2
particles in a constant magnetic field. Its Hamiltonian can be taken as
\begin{equation}\label{LMG}
  H=-\frac{\la}{N}\sum_{i<j}\left(\sigma^x_i\sigma^x_j+\gamma\sigma^y_i\sigma^y_j\right)
  -h\sum_i\sigma^z_i,
\end{equation}
where $\la>0$, $\ga\ge0$ and $h$ are real parameters, $\sigma^{a}_k$ is the $a$-th Pauli matrix
acting on the $k$-th spin, and the sums (as hereafter, unless otherwise stated) range from $1$ to
$N$. The model~\eqref{LMG} is thus the analogue of the Heisenberg~XY chain with long-range
constant interactions. Its Hamiltonian can be expressed in terms of the total spin operators
$S^a=\sum_i\sigma^a_i/2$ as
\begin{equation}\label{LMGSM}\fl
  H=-\frac{\la}{N}(1+\gamma)\Big(\mathbf{S}^2-(S^z)^2-\tfrac N2\Big)-2hS^z-
  \frac{\la}{2N}(1-\gamma)\Big((S^+)^2+(S^-)^2\Big),
\end{equation}
where $\mathbf{S}=(S^x,S^y,S^z)$ and $S^\pm=S^x\pm\iu S^y$.

It is well known~\cite{BJP82,BJ83} that the LMG model undergoes a second-order quantum phase
transition with mean field exponents at~$h/\la=1$. In the anisotropic case $0\le\ga<1$, the model
has been solved only in the thermodynamic limit~\cite{RVM07,RVM08}, although its entanglement
properties have been extensively studied (see, e.g.,~\cite{BDV06,ODV08,WVVD12}). In this work we
shall focus on the isotropic case~$\ga=1$, for which $H$ is diagonal in a basis $\ket{S,M,\nu}$ of
common eigenstates\footnote{The additional quantum number~$\nu$, which ranges from $1$ to
  $\binom{N}{N/2+S}-\binom{N}{N/2+S+1}$, takes into account the degeneracy of the eigenspace with
  a given~$S$ and~$M$.} of $\mathbf{S}^2$ and $S^z$ with respective eigenvalues~$S(S+1)$ and~$M$.
Here, as usual, $M=-S,-S+1,\dots, S$ and $S=\pi(N)/2,\pi(N)/2+1,\dots,N/2$, where $\pi(N)$ denotes
the parity of~the integer~$N$. We shall further set $\la=1$ (which amounts to a trivial rescaling)
and also restrict ourselves, without loss of generality, to nonnegative values of~$h$ (since the
spectrum of~$H$ is clearly independent of the sign of~$h$). With this choice of parameters
Eq.~\eqref{LMGSM} implies that~$H\ket{S,M,\nu}=E(S,M)\ket{S,M,\nu}$, where the eigenvalue~$E(S,M)$
is given by
\begin{eqnarray}
  E(S,M)
  &=-\frac{2}{N}\Big(S(S+1)-M^2-N/2\Big)-2hM\nonumber\\
  &=-\frac{2}{N}\,S(S+1)+\frac2N\left(M-\frac{Nh}2\right)^2
    -\frac{Nh^2}2+1\,.\label{eigvs}
\end{eqnarray}

The properties of the ground state of the LMG model~\eqref{LMG} with~$\la=\ga=1$ and $h\ge0$ are
essentially different in the two quantum phases~$0\le h<1$ and~$h\ge1$. Indeed, for~$h\ge1$ the
minimum of Eq.~\eqref{eigvs} is achieved when~$S=M=N/2$, so that the ground state is the product
state with all spins up. In particular, for $h\ge1$ the ground state is not entangled. On the
other hand, when $0\le h<1$ the energy is clearly a minimum when $S=N/2$ and $M=I(hN/2)$, where
$I(x)$ denotes the closest integer\footnote{The definition of~$I(x)$ is ambiguous when~$x$ is an
  integer (for odd $N$) or a half-integer (for even $N$). Since we are ultimately interested in
  the thermodynamic limit, we shall henceforth implicitly assume that $hN\ne 2p+1-\pi(N)$
  with~$p\in\ZZ$, so that $I(hN/2)$ is well defined.} (for~$N$ even) or half-integer (for $N$ odd)
to $x$. In particular, since the total spin~$S$ is maximum the ground state must be totally
symmetric. It is also non-degenerate, since the number~$N_\uparrow$ of ``up'' spins is fixed by
the condition
\begin{equation}\label{N0}\fl
  M=I(hN/2)=\frac12\,(N_\uparrow-N_\downarrow)=N_\uparrow-\frac N2\en\Longrightarrow\en N_\uparrow
  =\frac{N}2+I(hN/2)\,,
\end{equation}
where~$N_\downarrow=N-N_\uparrow$ denotes the number of ``down'' spins. Thus, when~$0\le h<1$ the
ground state of the LMG Hamiltonian~\eqref{LMG} with $\ga=\la=1$ is the totally symmetric
state\footnote{Note that in this case the quantum number $\nu$ can be omitted, since the subspace
  with $S=N/2$ and any $M$ is one-dimensional (cf.~the first footnote).}
\begin{equation}\label{GS2}
\ket{\psi(N_\uparrow,N_\downarrow)}\equiv\ket{N/2,I(hN/2)}\,,
\end{equation}
where $N_\uparrow$ is given by~\eqref{N0}. Denoting by $\ket{\!\!\uparrow}$ and
$\ket{\!\!\downarrow}$ respectively the spin up and down one-particle states, the (normalized)
ground state~\eqref{GS2} can be expressed in terms of the elements
\[
\ket{s_1,\dots,s_N}\equiv \ket{s_1}\cdots\ket{s_N}\equiv \ket{s_1}\otimes\cdots\otimes\ket{s_N}\,,
\qquad s_k=\uparrow,\downarrow\,,
\]
of the canonical spin basis as
\begin{equation}\label{LMGgs}
\ket{\psi(N_\uparrow,N_\downarrow)}=\binom{N}{N_\uparrow}^{-\frac12}\sum_{p\in
  S_{N_\uparrow,N_\downarrow}}
\ket{p(\underbrace{\uparrow,\dots,\uparrow,}_{N_\uparrow}
  \underbrace{\downarrow,\dots,\downarrow}_{N_\downarrow})}\,.
\end{equation}
In the latter formula $S_{N_\uparrow,N_\downarrow}$ denotes any set of~$\binom{N}{N_\uparrow}$
permutations of $N$ elements inequivalent with respect to the initial
state~$\ket{\!\!\uparrow,\dots,\uparrow,\downarrow,\dots,\downarrow}$ (i.e., such that the images
of the latter state under any two elements of the set differ).

Our aim is to construct a model generalizing the (isotropic) LMG model~\eqref{LMG} in two
different directions. More precisely, we shall consider an internal space of arbitrary dimension
$m+1$, and shall also allow for general (position-dependent) long-range interactions. We shall
only require that the ground state of the model (in the thermodynamic limit) be i)~non degenerate,
ii)~totally symmetric, and iii)~such that the number $N_s$ of particles in each one-particle state
$\ket s$ (with $s=1,\dots,m+1$) is well-defined, as in the original LMG model. In other words, the
ground state of the model should be the \emph{Dicke state}
\begin{eqnarray}
  \fl
  \ket{\psi(N_1,&\dots,N_{m+1})}\nonumber\\
  \fl&=\left(\frac{N!}{N_1!\cdots N_{m+1}!}\right)^{-\frac{1}{2}}
  \sum_{p\in
  S_{N_1,\dots,N_{m+1}}}|p(\underbrace{1,\dots,1}_{N_1},\ldots,\underbrace{m+1,\dots,m+1}_{N_{m+1}})\rangle,
  \label{gLMGgs}
\end{eqnarray}
with $N_1+\cdots+N_{m+1}=N$. As before, $S_{N_1,\dots,N_{m+1}}$ denotes any set of
$\frac{N!}{N_1!\cdots N_{m+1}!}$ permutations of $N$ elements inequivalent with respect to the
state $\ket{1,\dots,1,\dots,m+1,\dots,m+1}$.

In order to construct these generalized Lipkin--Meshkov--Glick (gLMG) models, we note that the
local $\su(2)$ spin operators~$\bS_k\equiv(S_k^x,S_k^y,S_k^z)$ are related to the spin permutation
operators~$S_{ij}$, whose action on the canonical spin basis is given by
\begin{equation}\label{Sij}
S_{ij}\ket{s_1,\dots,s_i,\dots,s_j,\dots,s_N}
=\ket{s_1,\dots,s_j,\dots,s_i,\dots,s_N}\,,
\end{equation}
by the identity
\[
\bS_i\cdot\bS_j=\frac12\left(S_{ij}-\frac12\right).
\]
From the previous equation it easily follows that
\[
\bS^2=\sum_{i<j}S_{ij}-\frac N4\,(N-4)\,,
\]
so that the Hamiltonian~\eqref{LMG} with $\la=\ga=1$ can also be  expressed as
\begin{equation}\label{HSij}
H=\frac2N\sum_{i<j}(1-S_{ij})+\frac2N\left(S^z-\frac{Nh}2\right)^2-\frac N2\,(1+h^2)\,.
\end{equation}
Motivated by this fact, we consider the general spin permutation operators $S_{ij}$ in
Eq.~\eqref{Sij} acting on particles with~$(m+1)$ internal degrees of freedom. It is well known
that these operators can be expressed in terms of the local (Hermitian) generators~$t^a_k$
($a=1,\dots,m(m+2)$) of the fundamental representation of the $\su(m+1)$ algebra acting on the
$k$-th site (with the normalization $\tr(t^a_kt^b_k)=\frac12\,\de_{ab}$) as
\begin{equation}\label{Sijta}
S_{ij}=\frac1{m+1}+2\sum_{a=1}^{m(m+2)}t^a_it^a_j\equiv\frac1{m+1}+2\,\bt_i\cdot\bt_j\,.
\end{equation}
Furthermore, since $S_{ij}$ is obviously Hermitian and has eigenvalues $\pm1$, it is clear that the
lowest energy eigenspace of the Hamiltonian
\begin{equation}
  H_0=\sum_{i<j}h_{ij}(1-S_{ij})
  \label{HSham}
\end{equation}
coincides with the subspace of totally symmetric states provided that $h_{ij}>0$ for all $i<j$.
Note that this Hamiltonian commutes with the total $\su(m+1)$ generators $T^a=\sum_{i}t^a_i$,
since these operators commute with each $S_{ij}$. In fact, the model~\eqref{HSham} reduces to the
well-known Haldane--Shastry spin chain of~$\su(m+1)$ type when the interactions are given by
\[
h_{ij}=\frac{\pi^2}{N^2}\,\sin^{-2}\left(\frac{\pi(i-j)}N\right)\,.
\]
This model, as well as its rational and hyperbolic versions, has been extensively studied in the
literature due to its remarkable integrability and solvability properties (see, e.g.,
Refs.~\cite{BGHP93, Po94, BFGR10}).

In order to single out a state of the form~\eqref{gLMGgs} as the unique ground state we need to
add a suitable term to the Hamiltonian $H_0$, as we shall now explain. By analogy with the
$\su(2)$ case, we define the operators~$J_i^a$ ($a=1,\dots,m$) acting on the $i$-th site
by
\begin{equation}\label{Jai}
J_i^a=E^{aa}_i-E^{m+1,m+1}_i\,,
\end{equation}
where~$E^{bc}$ denotes the $(m+1)\times(m+1)$ matrix whose only nonzero element is a~$1$ in the~$b$-th
row and $c$-th column\footnote{This particular choice of basis of the standard Cartan subalgebra
  of~$\su(m+1)$ is largely a matter of convenience, in that it results in the simplest form for
  Eqs.~\eqref{nas} and~\eqref{params} below. Note, in particular, that the operators~$J^a_i$ are
  not orthogonal with respect to the usual Killing--Cartan scalar product, i.e.,
  $\tr(J_i^aJ_i^b)\ne0$ for $a\ne b$.}. Note that the operators~$\iu J^a_k$ ($1\le a\le m$) are a
basis of the standard Cartan subalgebra of~$\su(m+1)$ (at the $k$-th site), and
that~$J^1_k=\si_k^z$ in the $\su(2)$ case. The simplest Hamiltonian with ground state satisfying
conditions~i)--iii) above is then given by
\begin{equation}\label{HgLMG}
  H = \sum_{i<j}h_{ij}(1-S_{ij})+\sum_{a=1}^mc_a(J^a-Nh_a)^2\equiv H_0+H_1\,,
\end{equation}
where $h_{ij},c_a>0$, $h_a\in\RR$ and
\[
J^a=\sum_iJ^a_i\,,\qquad 1\le a\le m\,.
\]
Indeed, note first of all that $H_0$ commutes with all the operators~$J^a$, and hence with $H_1$,
so that $H_0$ and $H_1$ can be simultaneously diagonalized. As mentioned before, the subspace of
totally symmetric states is the eigenspace of $H_0$ with lowest (zero) energy. On the other hand,
we have
\[\fl
J^a\ket{1,\dots,1,\dots,m+1,\dots,m+1}=(N_a-N_{m+1})\ket{1,\dots,1,\dots,m+1,\dots,m+1}\,,
\]
and therefore
\[
J^a\ket{\psi(N_1,\dots,N_{m+1})}=(N_a-N_{m+1})\ket{\psi(N_1,\dots,N_{m+1})}\,,
\]
since~$J^a$ commutes with all the permutation operators~$S_{ij}$. Thus in the thermodynamic limit
$N\to\infty$ with $N_a/N\to n_a$ finite, the energy of the Hamiltonian~\eqref{HgLMG} is minimum
for the state~$\ket{\psi(N_1,\dots,N_{m+1})}$ whose magnon densities~$n_a$ satisfy
\begin{equation}
  \label{Nasystem}
  n_a-n_{m+1}=h_a\,,\qquad 1\le a\le m\,.
\end{equation}
The above system is easily solved, with the result
\begin{equation}\label{nas}
n_a=h_a+\frac{1-h}{1+m}\,\qquad n_{m+1}=\frac{1-h}{1+m}\,,
\end{equation}
where
\[
h\equiv\sum_{a=1}^mh_a\,.
\]
The consistency condition for the system~\eqref{Nasystem}, namely that $0\le n_a\le 1$ for
$a=1,\dots,m+1$, is satisfied provided that
\begin{equation}\label{params}
-m\le h\le 1\,,\qquad -m\le h-(m+1)h_a\le 1\,.
\end{equation}
These are the equations of an $m$-dimensional (closed) simplex~$\cH\subset\RR^m$ with vertices
\begin{equation}\label{muas}
\bmu_a=\bev_a\quad (1\le a\le m)\,,\qquad\bmu_{m+1}=-\sum_{a=1}^m\bev_a\,,
\end{equation}
where $\{\bev_1,\dots,\bev_m\}$ is the canonical basis of~$\RR^m$ (cf.~Eq.~\eqref{Nasystem}). In
other words, the vectors $\bmu_a$ ($1\le a\le m+1$) are the weights of the fundamental
representation of~$\su(m+1)$ with respect to the basis~\eqref{Jai} of its Cartan subalgebra.

In summary, we have shown that (in the thermodynamic limit) the model~\eqref{HgLMG} with
$h_{ij},c_a>0$ and parameters $h_a$ satisfying~\eqref{params} has a non-degenerate, totally
symmetric ground state~$\ket{\psi(N_1,\dots,N_{m+1})}$ given by Eq.~\eqref{gLMGgs}, where
$N_a=n_aN$ and the magnon densities~$n_a$ are determined by $h_1,\dots,h_m$ through
Eq.~\eqref{nas}. The Hamiltonian~\eqref{HgLMG} is not invariant under the~$\su(m+1)$ algebra, due
to its second term~$H_1$. However, it can be easily expressed in terms of the (local) $\su(m+1)$
generators using Eq.~\eqref{Sijta}, namely
\[
H=-\sum_{i\ne j}h_{ij}\,\bt_i\cdot\bt_j+\sum_{a=1}^mc_a(J^a-Nh_a)^2+E_0\,,
\]
with $E_0=\frac m{m+1}\sum_{i<j}h_{ij}$. We shall thus refer to the model~\eqref{HgLMG} with $m+1$
internal degrees of freedom and $h_a$ satisfying Eq.~\eqref{params} as an~$\su(m+1)$ gLMG model.
Note that, since the parameters $h_a$ couple to the total $\su(m+1)$ generators~$J^a$, they can be
regarded as $\su(m+1)$ magnetic field strengths by analogy with the $\su(2)$ case. In fact, from
Eq.~\eqref{HSij} and the identity~$S^z=J^1/2$ it follows that the isotropic LMG model is
an~$\su(2)$ gLMG model with $h_{ij}=2/N$ for all $1\leq i,j\leq N$ and $c_1=1/(2N)$ (up to an
irrelevant constant).
\begin{rem}
  The condition that all the coefficients $h_{ij}$ in Eq.~\eqref{HgLMG} be positive can be
  considerably relaxed. Indeed, the latter condition is certainly sufficient to guarantee that the
  ground state of the model be symmetric, but it is by no means necessary. More precisely, it
  suffices that the transpositions~$i\leftrightarrow j$ corresponding to positive~$h_{ij}$
  generate the full symmetric group. For instance, the consecutive
  transpositions~$i\leftrightarrow i+1$ (with~$i=1,\dots,N-1$) certainly fulfill this requirement.
  It follows that the ground state of the Hamiltonian with nearest neighbor interactions
  \[
  H=\sum_{i}\tilde h_i(1-S_{i,i+1})+\sum_{a=1}^mc_a(J^a-Nh_a)^2\,,
  \]
  where $\tilde h_i,c_a>0$ and $S_{N,N+1}=1$ or $S_{N,N+1}=S_{1,N}$, is also of the
  form~\eqref{gLMGgs}. Thus the family of gLMG models can be suitably enlarged to encompass
  systems with both short-range and long-range interactions.
\end{rem}
\begin{rem}
  Although the ground state of the model~\eqref{HgLMG} with the restrictions~\eqref{params} is
  explicitly given by~Eq.~\eqref{gLMGgs}, its full spectrum cannot be computed in closed form for
  arbitrary values of the parameters~$h_{ij}$, $c_a$, $h_a$. Note, however, that the spin
  permutation operators~$S_{ij}$ leave invariant each of the subspaces~$\cS_{N_1,\dots,N_{m+1}}$
  consisting of states with~$N_s$ particles in each internal state $\ket s$ (magnons of type~$s$).
  It trivially follows that~$H_0$ leaves these subspaces invariant, and the same is true of~$H_1$,
  since each Cartan generator~$J^a$ is equal to the constant $N_a-N_{m+1}$ on them. Thus, in order
  to compute the spectrum of~$H$ it suffices to diagonalize the restrictions of $H_0$ to each of
  the subspaces~$\cS_{N_1,\dots,N_{m+1}}$. More precisely, if~$E_i(N_1,\dots,N_m)$ denotes an
  arbitrary eigenvalue of $H_0\big|_{\cS_{N_1,\dots,N_{m+1}}}$ then the eigenvalues of~$H$ are
  given by
  \[
  E_i(N_1,\dots,N_{m+1})+\sum_{a=1}^mc_a(N_a-N_{m+1}-Nh_a)^2\,.
  \]
\end{rem}

\section{Quantum entropies}\label{sec.entropies}

In this section we shall compute the analytic expressions of the (bipartite) entanglement
entropies of von Neumann and R\'enyi for the ground state of the gLMG model~\eqref{HgLMG} or, more
generally, of any quantum system whose ground state is a Dicke state~\eqref{gLMGgs}. Actually,
since the von Neumann entropy is the $q\to1$ limit of the R\'enyi entropy, it suffices to
compute~$\tr(\rho_L^q)$ for arbitrary~$q>0$. Note that, by Eq.~\eqref{rhoLLs}, the von Neumann
entropy is bounded above by~$\log d(L,m)$, where $d(L,m)$ is the dimension of the subspace spanned
by the Dicke states~$\ket{\psi(L_1,\dots,L_{m+1})}$ with $L_1+\cdots +L_{m+1}=L$, i.e., the
subspace of symmetric states for~$L$ particles with an~$(m+1)$-dimensional internal space. Since
we obviously have
\[
d(L,m)=\binom{L+m}{m}\le \frac{(L+m)^m}{m!}\,,
\]
the von Neumann entropy satisfies
\begin{eqnarray}
  \fl S&\le\log\binom{L+m}m\le m\log(L+m)-\log(m!)
         \le  m\log(L+m)-\int_1^m\log x\,\diff x\nonumber\\\fl
       &=m\log(L+m)-m\log m+m-1\,.\label{Sbound}
\end{eqnarray}
In other words, the von Neumann (bipartite) entropy of the ground state of the gLMG
model~\eqref{HgLMG} ---or, more generally, of any quantum system whose ground state is a Dicke
state~\eqref{gLMGgs}--- cannot grow faster than~$\log L$, which is the typical scaling behavior of
the entropy observed in many critical spin chains~(see, e.g.,\cite{LORV05}, \cite{ECP10}).

\medskip As first shown in Ref.~\cite{PSS05}, the reduced density matrix~$\rho_L$ is diagonal in
the Dicke basis~$\ket{\psi(L_1,\dots,L_{m+1})}$ (where $0\le L_i\le N_i$ and
$L_1+\cdots+L_{m+1}=L$), with eigenvalues
\begin{equation}
  \label{lambda}
  \la(L_1,\dots,L_m)=\binom NL^{\!\!-1}\prod_{a=1}^{m+1}\binom{N_a}{L_a}
\end{equation}
(see~\ref{app.rhoL} for a detailed derivation of the latter formula). Our first goal is to analyze
the behavior of $\la(L_1,\dots,L_m)$ in the thermodynamic limit $N\to\infty$, with
\begin{equation}\label{TD}
\lim_{N\to\infty}\frac LN=\al\,,\qquad \lim_{N\to\infty}\frac{N_a}N=n_a
\end{equation}
and $0<\al,n_a<1$ for $a=1,\dots,m+1$. Note that, by Eq.~\eqref{nas}, the latter condition on the
magnon densities~$n_a$ will be satisfied provided that the magnetic field strength
vector~$\bh\equiv(h_1,\dots,h_m)$ lies in the interior of the simplex~\eqref{params}. We start by
rewriting Eq.~\eqref{lambda} as
\begin{equation}\label{eigvhyp}
\la(L_1,\ldots,L_m)=\prod_{a=1}^m\la_a(L_1,\dots,L_a)\,,
\end{equation}
where each factor
\begin{equation}\label{laasdef}
\fl
\la_a(L_1,\ldots,L_a)=
{L-\sum\limits_{b<a} L_b\choose L_a}
{N-L-\sum\limits_{b<a}(N_b-L_b)\choose N_a-L_a}
{N-\sum\limits_{b<a}N_b\choose N_a}^{\!\!-1}
\end{equation}
is a hypergeometric distribution in the variable~$L_a$. Note, in particular, that all the binomial
coefficients appearing in the latter expression are well defined (i.e., non-vanishing) on account
of the inequalities~\eqref{Laineqs}. The main idea in order to derive the asymptotic behavior of
the RHS of Eq.~\eqref{eigvhyp} as $N\gg 1$ is to recursively apply the approximation of a
hypergeometric distribution by a suitable Gaussian distribution described in~\ref{app.hypgeom}. To
begin with, the first factor
\[
\la_1(L_1)=\binom L{L_1}\binom{N-L}{N_1-L_1}\binom N{N_1}^{-1}
\]
can be approximated using Eqs.~\eqref{plg}-\eqref{musigma} with
\[
\tilde L=L\,, \quad\tilde N=N\,,\quad l=L_1\,, \quad n=N_1\,,
\]
and hence $\tilde\al=\al$, $\nu=n_1$. We thus obtain (cf.~Eq.~\eqref{Gauss})
\[
\la_1(L_1)\simeq
g(L_1;\mu_1,\si_1)=\frac1{\sqrt{2\pi}\,\si_1}\,\e^{-\frac{x_1^2}{2\si_1^2}}\,,
\]
where
\[
\fl
\mu_1=Ln_1\,,\qquad \si_1^2=L(1-\al)n_1(1-n_1)\,,\qquad x_1=L_1-\mu_1=L_1-Ln_1\,.
\]
By Eqs.~\eqref{Liav} and~\eqref{xi2av}, the mean and standard deviation of the first magnon
number~$L_1$ are respectively equal to~$\mu_1$ and~$(1-1/N)^{-1}\si_1^2$, so
that~$\si_1/\mu_1=\Or(L^{-1/2})$. In particular, this implies that in the thermodynamic limit the
distribution of $L_1$ becomes sharply peaked around its mean value~$Ln_1$. As we shall now see,
this fact is crucial for determining the behavior of the second factor
\[
\la_2(L_1,L_2)=\binom {L-L_1}{L_2}\binom{N-L-N_1+L_1}{N_2-L_2}\binom {N-N_1}{N_2}^{-1}\,.
\]
Indeed, we can approximate~$\la_2(L_1,L_2)$ using Eqs.~\eqref{plg}-\eqref{musigma} with
\[
\tilde L=L-L_1\,,\quad \tilde N=N-N_1\simeq(1-n_1)N\,,\quad l=L_2\,, \quad n=N_2\,,
\]
and hence
\[
\tilde\al=\frac{\al-L_1/N}{1-n_1}\simeq\al\,,\qquad
\nu=\lim_{N\to\infty}\frac{N_2}{N-N_1}=\frac{n_2}{1-n_1}\equiv\nu_2\,,
\]
where we have used the fact that~$L_1\sim Ln_1$. We thus obtain
\[
\la_2(L_1,L_2) \simeq g(L_2;\mu_2,\si_2)\,,
\]
with
\begin{eqnarray*}
  \fl
  \mu_2&=\tilde L\nu_2=Ln_2-\frac{n_2x_1}{1-n_1}\,,\\\fl
  \si_2^2&=\tilde L(1-\tilde\al)\nu_2(1-\nu_2)\simeq L(1-n_1)(1-\al)\nu_2(1-\nu_2)=\frac{L(1-\al)n_2(1-n_1-n_2)}{1-n_1}\,,
\end{eqnarray*}
where in the second formula we have again taken into account that~$L_1\sim Ln_1$. Setting
$x_2=L_2-\mu_2$, from the previous formulas it immediately follows that
\[
\si_1\si_2=L(1-\al)n_1n_2(1-n_1-n_2)\,,\qquad L_2-\mu_2=x_2+\frac{n_2x_1}{1-n_1}\,,
\]
and hence
\[
\fl
\la_1(L_1)\la_2(L_1,L_2)=
\big[2\pi L(1-\al)\big]^{-1}\big[n_1n_2(1-n_1-n_2)\big]^{-1/2}
\exp\left(-\frac{\cE(x_1,x_2)}{2L(1-\al)}\right),
\]
where
\[
\fl
\cE(x_1,x_2)=\frac{x_1^2}{n_1(1-n_1)}
+\frac{1-n_1}{n_2(1-n_1-n_2)}\,\bigg(x_2+\frac{n_2x_1}{1-n_1}\bigg)^2=
\frac{x_1^2}{n_1}+\frac{x_2^2}{n_2}+\frac{(x_1+x_2)^2}{1-n_1-n_2}\,.
\]
The above approximate formula for $\la_1\la_2$ suggests that in general we have
\begin{equation}\label{laag}
  \fl
  \prod_{b=1}^a\la_b\simeq\big[2\pi L(1-\al)\big]^{-m/2}\bigg(1-\sum_{b=1}^an_b\bigg)^{-1/2}\prod_{b=1}^an_b^{-1/2}\exp\left(-\frac{\cE(x_1,\dots,x_a)}{2L(1-\al)}\right),
\end{equation}
with~$x_b\equiv L_b-Ln_b$ and
\begin{equation}
  \cE(x_1,\dots,x_a)=\sum_{b=1}^a\frac{x_b^2}{n_b}
  +\frac{\Big(\sum_{b=1}^ax_b\Big)^2}{1-\sum_{b=1}^an_b}\,.
  \label{sigmaa}
\end{equation}
This fact can be readily established by induction through a straightforward calculation. In
particular, setting $a=m$ in the previous formulas we obtain the following simple asymptotic
expression for the eigenvalues of the reduced density matrix~$\rho_L$ in the thermodynamic limit:
\begin{equation}
  \label{laasymp}
  \fl
  \la(L_1,\dots,L_m)\simeq\big[2\pi L(1-\al)\big]^{-m/2}\prod_{a=1}^{m+1}n_a^{-1/2}\cdot
  \exp\left(-\frac{\cE(x_1,\dots,x_m)}{2L(1-\al)}\right).
\end{equation}
\begin{rem}
  From the condition~$\tr\rho_L=1$ it follows that the integral over~$\RR^m$ of the RHS
  of~Eq.~\eqref{laasymp} is approximately equal to~$1$. In fact, it is straightforward to check
  that this integral is exactly equal to~$1$, which proves that the multivariate hypergeometric
  distribution~$\la(L_1,\dots,L_m)$ can be approximated in the thermodynamic limit~\eqref{TD} by a
  suitable normal distribution. We shall next show that this normal distribution is completely
  determined by the fact that its first and second moments coincide with those of the exact
  distribution~$\la(L_1,\dots,L_m)$ in the thermodynamic limit. In other words
  (cf.~Eqs.~\eqref{xi2avasymp} and~\eqref{xixjavasymp}), the RHS of Eq.~\eqref{laasymp} is simply
  the Gaussian distribution with moments
  \[
  \langle L_i\rangle=Ln_i\,,\qquad \big\langle L_iL_j\big\rangle- \big\langle L_i\big\rangle
  \big\langle L_j\big\rangle= \big\langle x_ix_j\big\rangle=L(1-\al)n_i(\de_{ij}-n_j)\,,
  \]
  where $x_i\equiv L_i-Ln_i$. Indeed, the covariance matrix of the general
  normal distribution
  \begin{equation}\label{gLas}
  g(L_1,\dots,L_m)=\frac{(\det A)^{1/2}}{(2\pi)^{m/2}}\exp\left(-\frac12\,\sum_{i,j=1}^ma_{ij}x_ix_j\right)
  \end{equation}
  with means~$\langle L_i\rangle=Ln_i$ is given by
  \[
  \big\langle x_ix_j\big\rangle=\frac{A_{ij}}{\det A}=(A^{-1})_{ij}\,,
  \]
  where~$A_{ij}$ is the complementary minor of~$a_{ij}$ in the symmetric
  matrix~$A\equiv(a_{ij})_{1\le i,j\le m}$. Comparing with Eq.~\eqref{gLas} we immediately obtain
  \[
  (A^{-1})_{ij}=L(1-\al)n_i(\de_{ij}-n_j)\,,
  \]
  which after an elementary calculation leads to
  \[
  a_{ij}=\big[L(1-\al)\big]^{-1}\left(\frac{\de_{ij}}{n_i}+\frac1{n_{m+1}}\right).
  \]
  This is exactly the coefficient matrix of the Gaussian distribution~\eqref{laasymp}, as claimed.
  In particular, this observation also shows that the approximation~\eqref{laasymp} coincides with
  that explicitly derived in Ref.~\cite{PSS05} for the case~$\al=0$ and~$m=2$.
\end{rem}

\medskip
Let us now turn to the computation of the trace
\[
\tr(\rho_L^q)=\sum_{L_1,\dots,L_m}\la(L_1,\dots,L_m)^q\,,
\]
where~$\la(L_1,\dots,L_m)$ is given by Eq.~\eqref{lambda} and the sum is over all non-negative
integers~$L_1,\dots,L_m$ satisfying the inequalities~\eqref{Laineqs}. In the thermodynamic limit
we can approximate~$\la(L_1,\dots,L_m)$ by the asymptotic formula~\eqref{laasymp}, and the sum
over~$L_1,\dots,L_m$ by an integral. Moreover, we can extend the domain of integration to the
whole space, since the Gaussian distribution is negligible unless $L_a\sim Ln_a$ for all~$a$. We
thus obtain
\begin{eqnarray}\fl
  \tr(\rho_L^q)
  &=\big[2\pi L(1-\al)\big]^{-mq/2}\prod_{a=1}^{m+1}n_a^{-q/2}\cdot\int_{\RR^m}
    \exp\left(-\frac{\cE(q^{1/2}x_1,\dots,q^{1/2}x_m)}{2L(1-\al)}\right)\diff x_1\cdots \diff x_m\nonumber\\\fl
  &=q^{-\frac m2}
    \big[2\pi L(1-\al)\big]^{-mq/2}\prod_{a=1}^{m+1}n_a^{-q/2}\cdot\int_{\RR^m}
    \exp\left(-\frac{\cE(x_1,\dots,x_m)}{2L(1-\al)}\right)\diff x_1\cdots \diff x_m\nonumber\\\fl
  &=q^{-\frac m2}\left(2\pi(1-\al)L\prod_{a=1}^{m+1} n_a^{1/m}\right)^{\!\!\frac{m(1-q)}2},
    \label{trrhoLq}
\end{eqnarray}
where we have used the fact that the RHS of~\eqref{laasymp} is normalized to~$1$.
\begin{rem}
  An immediate consequence of the Schmidt decomposition is the fact that
  $\tr(\rho^q_L)=\tr(\rho_{N-L}^q)$, where~$\rho_{N-L}$ is the reduced density matrix of the last
  $N-L$ spins~\cite{NC10}. It follows from this equality that~$\tr(\rho_L^q)$ should be symmetric
  under~$L\to N-L$. This obviously holds for the RHS of Eq.~\eqref{trrhoLq}, since
  \[
  L(1-\al)\simeq\frac{L(N-L)}N\,.
  \]
\end{rem}
\begin{rem}
  The~$\al\to0$ limit of Eq.~\eqref{trrhoLq} when~$q$ is a positive integer can also be obtained
  using the replica trick and the techniques developed in Refs.~\cite{CD11,CD12,CD13} to analyze
  totally symmetric states. More precisely, the authors of the latter references derive an
  expression for $\tr(\rho_L^q)$ (with $q\in\NN$) in the limit $N\to\infty$ with
  \begin{equation}\label{limCD}
    \lim_{N\to\infty}\frac{N_a}N=n_a\,,\qquad L~\text{fixed}\,,
  \end{equation}
  namely
  \[
  \tr(\rho_L^q)=\int_{[0,2\pi]^{mq}}
  \left(\sum_{a=1}^{m+1}n_a\e^{\iu(\th_a^{k+1}-\th_a^k)}\right)^{\!\!L}
  \prod_{k=1}^q\prod_{a=1}^m\frac{\diff\th^k_a}{2\pi}\,,
  \]
  where~$\th_a^{q+1}\equiv\th_a^1$ and~$\th_{m+1}^k=0$. Although this integral cannot be computed
  in closed form, its asymptotic behavior when $L\gg1$ can be exactly determined, with the result
  \begin{equation}\label{trCD}
    \tr(\rho_L^q)=q^{-\frac m2}\left(2\pi L\prod_{a=1}^{m+1}
      n_a^{1/m}\right)^{\!\!\frac{m(1-q)}2}\,.
  \end{equation}
  This is indeed the $\al\to0$ limit of Eq.~\eqref{trrhoLq}, as expected, since~\eqref{limCD}
  followed by the~$L\to\infty$ limit is essentially equivalent to the thermodynamic
  limit~\eqref{TD} followed by the $\al\to0$ limit. In any case, it should be noted that, although
  Eq.~\eqref{trCD} can be obtained from~\eqref{trrhoLq} taking the~$\al\to0$ limit (at least
  for~$q\in\NN$), it is of course not possible to derive the latter equation from~\eqref{trCD}.
\end{rem}

\medskip From Eqs.~\eqref{VN-Renyi} and~\eqref{trrhoLq} it immediately follows that in the large
$L$~limit the R\'enyi entropy is given by
\begin{equation}\label{Rfinal}
R_q=\frac{m}{2}\frac{\log q}{q-1}+\frac{m}{2}\log\left(2\pi L(1-\alpha)
  \prod_{a=1}^{m+1} n_a^{1/m}\right).
\end{equation}
Remarkably, $R_q$ depends on~$q$ only through the first term, which is irrelevant in the
thermodynamic limit~$L\to\infty$. This fact, already noted by the authors of Ref.~\cite{CD13} in
the $\al=0$ case, shall prove important for determining whether the class of generalized LMG
models~\eqref{HgLMG} contains critical models. Taking the~$q\to1$ limit of Eq.~\eqref{Rfinal} we
deduce a similar formula for the von Neumann entropy, namely
\begin{equation}\label{SvN}
  S
  =\frac{m}{2}\log\left(2\pi\e L(1-\alpha)
    \prod_{a=1}^{m+1} n_a^{1/m}\right).
\end{equation}
Equation~\eqref{SvN} was first derived in Ref.~\cite{PSS05} by extrapolation from the~$\al=0$ case
(see also Ref.~\cite{PS05}). Note also that this equation is clearly consistent with the a priori
bound~\eqref{Sbound} since~$L\gg1$ by hypothesis.

\begin{figure}[h]
  \centering
  \includegraphics[width=9.5cm]{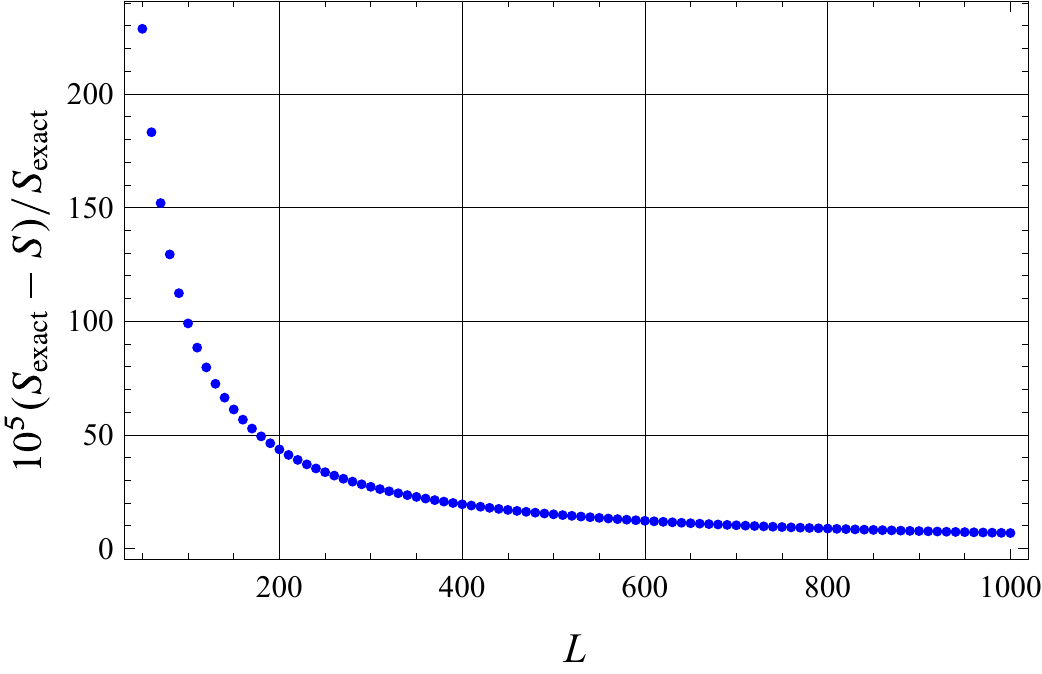}
  \caption{Relative error of the approximation~\eqref{SvN} to the exact von Neumann
    entropy~$S_{\mathrm{exact}}$ for $m=2$, $h_1=h_2=1/5$, $\al=1/2$ as a function of
    $L\in[50,1000]$ (in increments of~$10$).}
  \label{fig.relerr}
\end{figure}

Equations~\eqref{Rfinal}-\eqref{SvN} provide excellent approximations to the exact values of the
R\'enyi and von Neumann entropies for even moderately large values of~$L$, with a relative error
steadily decreasing with~$L$ (see, e.g., Fig.~\ref{fig.relerr} for the case~$m=2$, $h_1=h_2=1/5$
and~$\al=1/2$). Remarkably, in this case the R\'enyi and von Neumann entropies merely differ by a
($q$-dependent) constant, namely
\[
R_q=S+\frac m2\left(\frac{\log q}{q-1}-1\right).
\]
Note also that in the $\su(2)$ case the von Neumann entropy~\eqref{SvN} reduces to
\[
S=\frac12\log\Bigl(2\pi\e L(1-\al)n_1n_2\Bigr)=\frac12\log\bigg[\frac{\pi\e}2
L(1-\al)\bigg]+\frac12\log(1-h^2)\,,
\]
in agreement with the result of Refs.~\cite{PS05} and~\cite{LORV05} for the isotropic LMG model.

As expected, the maximum value of the entropies~\eqref{Rfinal}-\eqref{SvN} is obtained when
$n_a=1/(m+1)$ for all $a=1,\dots,m+1$, i.e., for~$\bh=0$. On the other hand, the
approximations~\eqref{Rfinal}--\eqref{SvN} to the von Neumann and R\'enyi entropies tend
to~$-\infty$ when~$\bh$ approaches the faces of the simplex~$\cH$, since each of these faces is
determined by the vanishing of one of the magnon densities~$n_a$. This behavior had already been
pointed out in Ref.~\cite{LORV05} for the von Neumann entropy of the isotropic ($\su(2)$) LMG
model. It may seem surprising that the approximate von Neumann and R\'enyi
entropies~\eqref{Rfinal}--\eqref{SvN} become negative in a certain subset of~$\cH$. In fact, as we
shall now discuss in more detail, when $L\gg1$ the regions in which each of these approximations
are negative (and, therefore, break down) are negligibly small.
  
In order to substantiate the previous claim, we shall estimate the distance $r_0$ of a point
$\bh_0$ lying on one of the $(m-1)$-dimensional faces of the simplex~$\cH$ to the positive entropy
region, or equivalently to the zero entropy hypersurface. For simplicity, we shall deal with the
generic situation in which~$\bh_0$ belongs to the interior of the face. We shall prove that for
both entropies under consideration $r_0\sim L^{-m}$. To this end, suppose to begin with
that~$\bh_0$ lies in the interior of the face where~$n_a=0$, with~$1\le a \le m$. In this case we
can approximate the remaining densities~$n_b$ by their values $n_b^0=h_{0b}+(1-h_0)/(m+1)$ at the
point~$\bh_0\equiv(h_{01},\dots,h_{0m})$ (with $h_0\equiv\sum_{a=1}^{m+1}h_{0a}$), so that (for
instance) the von Neumann entropy satisfies
\[
S\simeq \frac m2\,\log\big[2\pi\e L(1-\al)\big]+\frac12\hps{1\le b} \sum_{\mathclap{1\le b\ne a\le
    m+1}}\,\log n_b^0+\frac12\,\log n_a\,.
\]
Equating the RHS of this equation to~zero and solving for~$n_a$ we obtain
\[
n_a\equiv \frac1{m+1}\,\left(mh_a-\sum_{\mathclap{1\le b\ne a\le m+1}}h_b+1\right)=
\frac{\big[2\pi\e L(1-\al)]^{-m}}{\prod\limits_{\mathclap{1\le b\ne a\le m+1}}n_b^0}\,.
\]
Thus near~$\bh_0$ the $S=0$ hypersurface can be approximated by the above hyperplane, which is
parallel to the hyperplane $n_a=0$ containing the face under consideration. Computing the distance
between these two hyperplanes we immediately find the following approximate formula for~$r_0$ in
the case of the von Neumann entropy:
\begin{equation}\label{r0vN}
  r_0\simeq \frac{m+1}{\sqrt{m^2+m-1}}\,\frac{\big[2\pi\e
    L(1-\al)\big]^{-m}}{\prod\limits_{\mathclap{1\le b\ne a\le m+1}}n_b^0}\sim L^{-m}\,.
\end{equation}
When~$\bh_0$ lies on the face~$n_{m+1}=0$ a totally analogous calculation leads to the slightly
simpler result
\begin{equation}\label{r0vNlast}
  r_0\simeq \frac{m+1}{\sqrt{m}}\,\frac{\big[2\pi\e
    L(1-\al)\big]^{-m}}{\prod\limits_{\mathclap{1\le b\le m}}n_b^0}\sim L^{-m}\,.
\end{equation}
The computation of $r_0$ for the R\'enyi entropy proceeds along the same lines, with the result
\begin{equation}\label{r0RT}
  r_0\simeq q^{\frac m{1-q}}\,\e^{m}\,(r_0)_{\mathrm S}\,,
\end{equation}
where~$(r_0)_{\mathrm{S}}$ is the approximate value of~$r_0$ for the von Neumann entropy given by
Eqs.~\eqref{r0vN} or \eqref{r0vNlast}. Interestingly, the RHS of Eq.~\eqref{r0RT} is less
(resp.~greater) than~$(r_0)_{\mathrm S}$ for~$0<q<1$ (resp.~$q>1$). Note, finally, that the
assumption that~$\bh_0$ belongs to the \emph{interior} of the face is essential for the validity
of Eqs.~\eqref{r0vN}--\eqref{r0RT}. More generally, it can be shown that the corresponding value
of~$r_0$ for~$\bh_0$ lying on (the interior of) an~$(m-k)$-dimensional face of~$\cH$ (with
$k=1,\dots,m$) behaves as~$L^{-m/k}$. In other words, even in the most unfavorable case~$k=m$
(i.e., when~$\bh_0$ is one of the vertices of the simplex~$\cH$) the distance of~$\bh_0$ to the
positive entropy region is $\Or(L^{-1})$.

\medskip By Eqs.~\eqref{Rfinal}-\eqref{SvN}, when~$L\to\infty$ the von Neumann and R\'enyi ground
state entanglement entropies
satisfy
\begin{eqnarray}\label{RnNlim}
  S=R_q&=\frac{m}{2}\log L+\Or(1)\,.
       \label{Tlimi}
\end{eqnarray}
As remarked above, this behavior of the von Neumann entropy is characteristic of quantum critical
(gapless) one-dimensional lattice systems with short-range interactions. More generally,
when~$d>1$ the von Neumann entanglement entropy of a $d$-dimensional critical system featuring
short-range interactions is expected to scale as~$L^{d-1}\log L$ (for fermionic systems) or
$L^{d-1}$ (for bosonic ones) when~$L\gg1$, $L$ being the linear size of the system. On the other
hand, for a non-critical (gapped) system with short-range interactions the von Neumann entropy
should grow only as~$L^{d-1}$. This so-called \emph{area law}~\cite{ECP10} has been verified for a
wide range of quantum systems, such as the XX and XY models~\cite{JK04,IJK05}, the Heisenberg
(XYZ) spin chain~\cite{LRV04,EER10}, the original ($\su(2)$, not necessarily isotropic) LMG
model~\cite{LORV05}, translation-invariant (quadratic) fermionic systems in arbitrary
dimension~\cite{Wo06}, and certain two-dimensional bosonic and fermionic systems~\cite{CT08},
\cite{BCS06}, to name only a few. On the other hand, for models with long-range interactions it is
widely accepted that the area law need not hold, since in general the range of the interaction
tends to increase the entropy~\cite{ECP10}. This statement should be taken with some caution,
since the entanglement entropy is ultimately a property of the state, and two models featuring
short-range and long-range interactions may have the same ground state~\cite{CSWCP13}. The
logarithmic growth~\eqref{Tlimi} of the ground state entanglement entropy of the
gLMG~models~\eqref{HgLMG} does \emph{not} indicate, however, that this class contains critical
models. Indeed, the R\'enyi entropy of these models scales as~$a\log L$ with~$a$ independent of~$q$,
while for a two-dimensional CFT the coefficient~$a$ should instead be proportional to $1+q^{-1}$.

It has been shown in Ref.~\cite{CT08} that in some low-dimensional quantum systems whose
entanglement (von Neumann) entropy follows the area law the Tsallis entropy, defined by
\begin{equation}
  T_q=\frac{\tr(\rho_L^q)-1}{1-q}\,,
  \label{Tsallis}
\end{equation}
becomes extensive (i.e., scales as $L^d$, where~$d$ is the number of space dimensions and~$L$ is a
characteristic length) for a suitable value of the positive parameter~$q$. This entropy, which
plays an important role in the study of strongly correlated classical systems, has also been
extensively applied in other fields ranging from natural and social sciences to linguistics and
economics (see the online document \texttt{http://tsallis.cat.cbpf.br/TEMUCO.pdf} for an updated
bibliography). Since the Tsallis and R\'enyi entropies are obviously related by
\begin{equation}\label{RqTq}
  R_q=\frac{\log\bigl[1+(1-q)T_q\bigr]}{1-q}\,,
\end{equation}
from Eq.~\eqref{Rfinal} we immediately obtain the following explicit formula for the Tsallis
entanglement entropy of the ground state~\eqref{gLMGgs} of the gLMG model~\eqref{HgLMG}:
\begin{equation}
  T_q= \frac{1}{1-q}\left(
    q^{-\frac m2}\left[2\pi L(1-\alpha)
      \prod_{a=1}^{m+1} n_a^{1/m}
    \right]^{\frac{m(1-q)}2}-1\right).
  \label{Tfinal}
\end{equation}
Thus for $L\gg1$ the Tsallis entropy scales as a power law, namely
\begin{equation}\label{TqLgg1}
  T_q= 
  \frac{L^{\frac{m(1-q)}2}}{(1-q)q^{\frac m2}}\left[2\pi (1-\alpha)
    \prod_{a=1}^{m+1} n_a^{1/m}
  \right]^{\frac{m(1-q)}2}+\Or(1).
\end{equation}
The dominant term of~$T_q$ is linear in $L$ if $q=1-2/m$, which requires that $m\ge3$ due to the
condition~$q>0$. For this critical value of $q$ the Tsallis entropy is given by
\begin{equation}\label{Tcrit}
T_{1-\frac2m}= 
\frac{\pi m L(1-\alpha)}{{(1-\frac2m)}^{\frac m2}}
\prod_{a=1}^{m+1} n_a^{1/m}
 -\frac m2\,,
\end{equation}
so that the entropy per particle in the thermodynamic limit reads
\begin{equation}\label{Tcritlim}
  \lim_{L\to\infty}\frac{T_{1-\frac2m}}{L}= 
  \frac{\pi m(1-\alpha)}{{(1-\frac2m)}^{\frac m2}}
  \prod_{a=1}^{m+1} n_a^{1/m}\,.
\end{equation}
As we have just seen, in the~$\su(2)$ and $\su(3)$ cases the Tsallis entanglement entropy is not
extensive for \emph{any} value of the parameter~$q$. It would therefore be of interest, in this
context, to find a generalized entropy (like, e.g., one of the group entropies studied in
Refs.~\cite{Te15} and~\cite{Te16}) which is extensive for the ground state of the gLMG
model~\eqref{HgLMG} with~$m=1,2$.

\section{Ground state phase diagram}\label{sec.gspd}

In the previous sections we have studied the entanglement properties of the ground state of the
gLMG model~\eqref{HgLMG} when the magnetic field strength $\bh$ lies in the interior of the
simplex~$\cH$ given by Eq.~\eqref{params}. The aim of this section is to extend the previous
results outside this region, determining how the behavior of the ground state and its entanglement
entropy vary with~$\bh$.

To this end, note first of all that by Eq.~\eqref{HgLMG} the magnon densities~$n_a$ in the ground
state must minimize the function
\begin{equation}\label{vepns}
\vep(n_1,\dots,n_m)=\sum_{a=1}^mc_a(n_a-n_{m+1}-h_a)^2\,,
\end{equation}
with $n_{m+1}=1-\sum\limits_{a=1}^mn_a$, in the simplex~$\cN\subset\RR^m$ defined by
\[
0\le n_a\le 1\,,\qquad \sum_{a=1}^mn_a\le 1\,.
\]
Thus the condition for the ground state to have well-defined magnon densities is that~$\vep$ has a
\emph{unique} minimum in~$\cN$. In fact, from the very definition of the set~$\cH$ it follows that
for~$\bh\in\cH$ the unique minimum of the energy function~\eqref{vepns} over~$\RR^m$, given by
Eqs.~\eqref{Nasystem}, lies in~$\cN$. More precisely, when~$\bh$ belongs to the interior of~$\cH$
the absolute minimum of~$\vep$ lies in~the interior of~$\cN$, and therefore $0<n_a<1$ for
all~$a=1,\dots,m+1$. Thus in this case the ground state is not only entangled, but contains
magnons of each of the~$m+1$ types~$\ket a$. On the other hand, when~$\bh$ belongs to the boundary
of~$\cH$ the unique global minimum of~$\vep$, still given by Eq.~\eqref{nas}, now lies on the
boundary of~$\cN$. In particular, at least one of the magnon densities must vanish in this case,
so that the ground state is entangled but does not have full magnon content. We shall be mainly
interested in this section in the case in which $\bh$ lies in the exterior of~$\cH$, so that the
minimum of~$\vep$ in~$\cN$ is necessarily attained on its boundary~$\pd\cN$. This minimum is
therefore no longer given by the simple equations~\eqref{Nasystem}, but must be computed by
examining the behavior of~$\vep$ on~$\pd\cN$. We shall prove at the end of this section that the
minimum of~$\vep$ on~$\pd\cN$ is unique for all values of the magnetic field strength~$\bh$. Thus
the ground state of the generalized LMG model~\eqref{HgLMG} is always unique, although its magnon
content is given by Eq.~\eqref{nas} only for~$\bh\in\cH$.

The boundary of the simplex~$\cN$ is the union of the~$(m-k)$-dimensional sets (or
$(m-k)$-\emph{faces}) $\pd\cN_{a_1,\dots,a_k}$ determined by the equations
\[
n_{a_1}=\cdots=n_{a_k}=0\,,
\]
where~$k=1,\dots,m$ and $1\le a_1<\cdots<a_k\le m+1$. Let us suppose, therefore, that the unique
minimum of~$\vep$ in~$\pd\cN$ is attained on a certain $(m-k)$-face~$\pd\cN_{a_1,\dots,a_k}$.
If~$k=m$ (i.e., if the minimum of~$\vep$ on~$\cN$ is attained at one of its vertices) then one of
the magnon densities is necessarily equal to $1$ and the remaining ones vanish, so that the ground
state is not entangled. On the other hand, if~$1\le k<m$ then the ground state is entangled but
has not full magnon content, since it only contains magnons of $m-k+1<m+1$ types. In other words,
we expect that in general the ground state can be in exactly one of~$m+1$ possible ``phases'',
characterized by the vanishing of $0\le k\le m$ magnon densities. Note that we have included
the~$k=0$ case, in which $n_a>0$ for all~$a=1,\dots,m+1$, which holds when~$\bh$ lies in the
interior of~$\cH$.

We shall now describe the essential properties of the ground state when~$\bh$ belongs to the
exterior of~$\cH$. As we have just seen, in this case the minimum of the energy function~$\vep$ is
attained on one of the $(m-k)$-faces~$\pd\cN_{a_1,\dots,a_k}$
of~$\cN$, so that in the ground state
$n_{a_1}=\cdots=n_{a_k}=0$ and $0<n_{a_{k+1}},\dots,n_{a_{m+1}}<1$, where
\[
\{a_{k+1},\dots,a_{m+1}\}=\{1,\dots,m+1\}\setminus\{a_{1},\dots,a_{k}\}\,.
\]
Thus in the thermodynamic limit~\eqref{TD} we have
\[
N_{a_1}=\cdots=N_{a_k}=L_{a_1}=\cdots =L_{a_k}=0\,,
\]
and the eigenvalues of the reduced density matrix~$\rho_L$ are given by
\begin{eqnarray*}
  \la(L_{a_{k+1}},\dots,L_{a_{m}})
  &=\frac{L!}{\prod_{j=k+1}^{m+1}
  L_{a_j}!}\frac{(N-L)!}{\prod_{j=k+1}^{m+1}(N_{a_j}-L_{a_j})!}
    \frac{\prod_{j=k+1}^{m+1}N_{a_j}!}{N!}\\
  &=\binom NL^{\!\!-1}\prod_{j=k+1}^{m+1}\binom{N_{a_j}}{L_{a_j}}\,,
\end{eqnarray*}
with $L_{a_{m+1}}=L-\sum_{j=k+1}^mL_{a_j}$\,. The distribution of the $m-k$ independent
variables~$L_{a_j}$ ($k+1\le j\le m$) is therefore the analogue of Eq.~\eqref{lambda}, with $m$
replaced by~$m-k$ and~$n_1,\dots,n_{m+1}$ by~$n_{a_{k+1}},\dots,n_{a_{m+1}}$. Hence
Eqs.~\eqref{Rfinal}, \eqref{SvN} and~\eqref{Tfinal} for the R\'enyi, von Neumann and Tsallis
entropies still hold in this case, provided that we perform the above replacements; in particular,
the von Neumann entropy is given by
\begin{equation}\label{SvNk}
S=\frac{m-k}2\log\left(2\pi\e L(1-\alpha)
    \prod_{j=k+1}^{m+1} n_{a_j}^{1/(m-k)}\right)\,.
\end{equation}
In the derivation of the latter equation we have tacitly assumed that $k<m$. In fact, when~$k=m$
the ground state is not entangled, and hence~$S=0$ in this case.

By Eq.~\eqref{SvNk}, the asymptotic behavior of the von Neumann entropy in the ground state phase
with~$k$ vanishing magnon densities is given by
\[
S=\frac{m-k}2\,\log L+\Or(1)\,.
\]
As discussed in the previous section, this result does not imply that in this phase there should
be critical gLMG models. Indeed, the R\'enyi entropy $R_q$ also scales as $a\log L$ when~$L\gg1$,
where~$a=(m-k)/2$ is independent of the parameter~$q$. As we know, this behavior is inconsistent
with the characteristic scaling of the R\'enyi entropy of a one-dimensional CFT, for which
$R_q=a(q)\log L$ with $a(q)$ proportional to~$1+q^{-1}$.
\begin{rem}
  By the discussion preceding Eq.~\eqref{SvNk}, the Tsallis entropy in the phase with $k$
  vanishing magnon densities scales as~$L^{(m-k)(1-q)/2}$, and is therefore extensive for
  $q=1-2/(m-k)$ provided that~$m-k\ge3$. Hence this entropy is not extensive for \emph{any} value
  of $q$ in the phases with $m-1$ and $m-2$ vanishing magnon densities.
\end{rem}

As a concrete example of the previous general statements, we shall next discuss in detail
the~$\su(3)$ case ($m=2$) with the symmetric choice $c_1=c_2=C$. Taking (without loss of
generality) $C=1$, the energy function is simply
\begin{equation}
  \label{enfunvec}
  \vep(\bn)=\big(\bx(\bn)-\bh\big)^2\,,
\end{equation}
where $\bx(\bn)\equiv\sum_{s=1}^3n_s\bmu_s$, $\bn=(n_1,n_2)\in\cN$ and $n_3=1-n_1-n_2$. When~$\bn$
ranges over~$\cN$, the point~$\bx(\bn)$ varies over the triangle~$\cH$ in a one-to-one fashion,
with~$\bn\in\pd\cN$ if and only if~$\bx(\bn)\in\pd\cH$. Hence the minimum of~$\vep(\bn)$ is simply
the distance of the fixed vector~$\bh$ to the triangle~$\cH$. Moreover, since this triangle is
convex, the minimum distance of~$\bh$ to~$\cH$ is attained at a unique point~$\bx(\bn_0)$
in~$\cH$.

\begin{figure}[h]
  \centering
  \includegraphics[height=7cm]{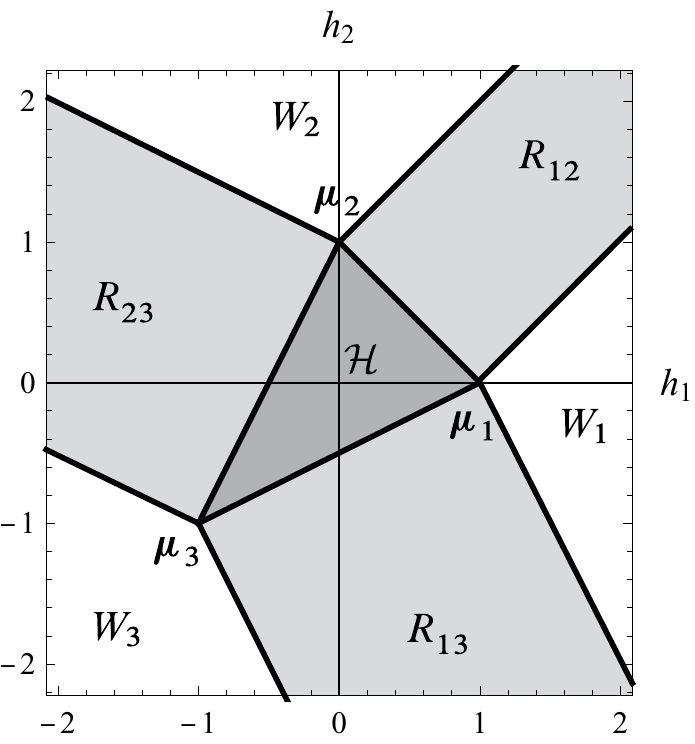}\kern.9em\hfill
  \caption{Triangle~$\cH$, half-strips~$R_{ab}$, wedges~$W_a$ and vertices~$\bmu_a$ in
    the~$(h_1,h_2)$-plane.}
  \label{fig.Abc}
\end{figure}

\begin{figure}[h]
  \centering
  \includegraphics[width=.6\textwidth]{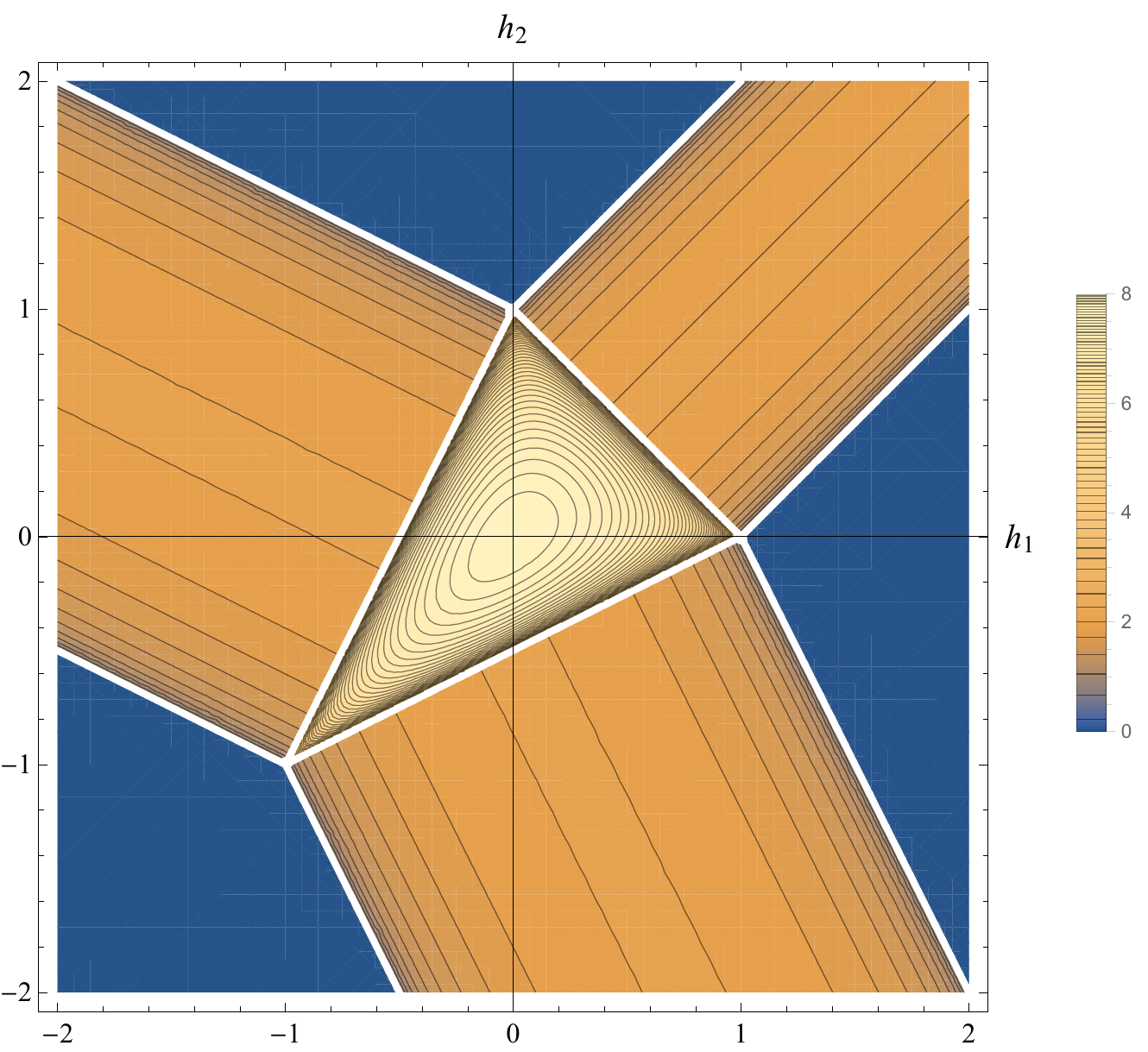}
  \caption{Von Neumann entropy of the~$\su(3)$ gLMG model~\eqref{HgLMG} with~$c_1=c_2$ and
    $L(1-\al)=1000$ as a function of the magnetic field~$\bh=(h_1,h_2)$.}
  \label{fig.splot}
\end{figure}

The problem of minimizing the energy function~$\vep$ has a very simple geometric solution. Indeed,
when~$\bh\in\Int\cH$ the point~$\bx(\bn_0)$ closest to~$\bh$ is obviously~$\bh$ itself, so that
the magnon density vector~$\bn_0=\bx^{-1}(\bh)$ is given by Eq.~\eqref{nas}. Suppose, on the other
hand, that~$\bh\notin\Int\cH$ (including the limiting case~$\bh\in\pd\cH$). Let us denote
by~$L_{bc}$ the side of the triangle~$\cH$ with vertices~$\bmu_b$ and $\bmu_c$. This side is
parametrized by a magnon density~$\bn$ with~$n_a=0$, where~$\{a,b,c\}=\{1,2,3\}$. Likewise, we
shall denote by $R_{bc}$ the half-strip bounded by the side~$L_{bc}$ and the two straight lines
perpendicular to it through the vertices~$\bmu_b$ and~$\bmu_c$ which does not contain the opposite
vertex~$\bmu_a$, with the two limiting half-lines removed (cf.~Fig.~\ref{fig.Abc}). By
construction, when~$\bh\in R_{bc}$ the point in~$\cH$ closest to~$\bh$ lies on the interior of the
side~$L_{bc}\subset\pd\cH$, so that in this case the corresponding density~$\bn_0$ belongs to the
side~$n_a=0$ of~$\cN$. The value of~$\bn_0\equiv(n_{01},n_{02})$ can be easily computed by
minimizing
\begin{equation}\label{vepna0}
\vep\big|_{n_a=0}=\Big(n_b(\bmu_b-\bmu_c)+\bmu_c-\bh\Big)^{\!2}
\end{equation}
with respect to~$n_b$ (assuming, without loss of generality, that~$b\in\{1,2\}$). In this way we
easily obtain
\begin{equation}\label{nbLbc}
  n_{0b}=\frac{(\bh-\bmu_c)\cdot(\bmu_b-\bmu_c)}{(\bmu_b-\bmu_c)^2}\,,
\end{equation}
and~of course~$n_{0c}=1-n_{0b}$. Note that in this case, although~$n_{0a}=0$,~the ground state is
still entangled, since~$0<n_{0b},n_{0c}<1$. On the other hand, let~$W_a$ denote the closed wedge
with vertex~$\bmu_a$ limited by the half-strips~$R_{ab}$ and~$R_{ac}$ (cf.~Fig.~\ref{fig.Abc}).
Obviously, if~$\bh$ lies in $W_a$ the point of the triangle~$\cH$ closest to~$\bh$ is the
vertex~$\bmu_a$, whose corresponding magnon densities are~$n_a=1$ and~$n_b=n_c=0$.
In summary, we have shown that the ``phase diagram'' of the ground state of the~$\su(3)$ gLMG
model~\eqref{HgLMG} (with~$c_1=c_2$ for all $a$) is as follows:
\begin{enumerate}[i)]
\item In the interior~$\Int\cH$ of the triangle~$\cH$, the ground state is a symmetric state
  containing all three types of magnons~$\ket a$ (with~$a=1,2,3$).
\item In each of the sets~$R_{bc}$ the ground state is still entangled, but contains only magnons
  of the two types~$\ket b$ and~$\ket c$.
\item In the wedges~$W_a$, the ground state consists of magnons of type~$\ket a$ only, and is
  therefore not entangled.
\end{enumerate}
From the previous remark, the general formula~\eqref{SvNk} and Eqs.~\eqref{nas} and~\eqref{nbLbc},
it follows that when~$L\gg1$ the von Neumann entanglement entropy as a function of the~$\su(3)$
magnetic field strength~$\bh$ is given by
\[
\fl S= \cases{
  S_0-\frac32\log3+\frac12\,\Big(\log(1+2h_1-h_2)+\log(1-h_1+2h_2)&\vspace*{-.5\baselineskip}\\
  \hphantom{S_0-\frac32\log3+\frac12\,\big(\log(2h_1-h_2-1)+\log(2h_2-}+\log(1-h_1-h_2)\Big),
  &$\bh\in\Int\cH$\\
  \frac12\Big(S_0+\log\left[1-(h_1-h_2)^2\right]-2\log2\Big),&
  \hphantom{$\bh\in\Int\cH$}\llap{$\bh\in R_{12}$}
  \\
  \frac12\Big(S_0-2\log 5+\log(3+2h_1+h_2)+\log(2-2h_1-h_2)\Big),&
  \hphantom{$\bh\in\Int\cH$}\llap{$\bh\in R_{13}$}\\
  \frac12\Big(S_0-2\log 5+\log(3+h_1+2h_2)+\log(2-h_1-2h_2)\Big),&
  \hphantom{$\bh\in\Int\cH$}\llap{$\bh\in R_{23}$}\\
  0\,,&\hphantom{$\bh\in\Int\cH$}\llap{$\bh\in W_1\cup W_2\cup W_3\,,$}
}
\]
with $S_0\equiv\log\bigl[2\pi\e L(1-\al)\bigr]$\,. In Fig.~\ref{fig.splot} we have plotted this
entropy as a function of the magnetic field~$\bh$ in the range~$|h_a|\le 2$ for $L(1-\al)=1000$.
As mentioned in the previous section, the latter approximation to the von Neumann entropy tends
to~$-\infty$ when~$\bh$ approaches a side~$L_{ab}$ of the triangle~$\cH$ from its interior. Note,
however, that it has a finite limit when this side is approached from the corresponding
half-strip~$R_{ab}$. Similarly,~$S\to-\infty$ when~$\bh$ approaches one of the straight lines
limiting a half-strip~$R_{ab}$ from its interior, but has a finite limit when this straight line
is approached from the corresponding wedge~$W_a$ or $W_b$ (except at the vertices~$\bmu_a$
and~$\bmu_b$).
\begin{rem}
  A formula similar to the previous equation for the von Neumann entropy can be derived without
  difficulty for the R\'enyi and the Tsallis entropies. In fact, comparing
  Eqs.~\eqref{Rfinal}-\eqref{SvN} it is clear that the only difference between the von Neumann and
  the R\'enyi entropies is the constant term $(m-k)[(\log q)/(q-1)-1]/2$ in the phase with~$k$
  vanishing magnon densities (i.e., $k=0,1,2$ respectively for $\bh$ belonging to~$\Int\cH$,
  $R_{ab}$ and $W_a$). Thus the von Neumann and R\'enyi entropies are essentially equivalent for the
  models under consideration.
\end{rem}

\medskip
In the general case (i.e., for $m>2$ and arbitrary positive values of the parameters~$c_a$), the
analysis is very similar. Indeed, in this case the energy function~$\vep(\bn)$ can be written as
\[
\vep(\bn)=\Vert\bx(\bn)-\bh\Vert^2\,,
\]
where
\[
\bx(\bn)=\sum_{s=1}^{m+1}n_s\bmu_s\,,\qquad \bn=(n_1,\dots,n_m)\,,\qquad
n_{m+1}=1-\sum_{a=1}^mn_a\,,
\]
and the norm~$\Vert{\cdot}\Vert$ is defined by~$\Vert\by\Vert^2\equiv\sum_{a=1}^mc_ay_a^2$. As
before, when the vector~$\bn$ varies over the simplex~$\cN$ the point~$\bx(\bn)$ parametrizes the
simplex~$\cH$ in a bijective way, with~$\bn\in\pd\cN$ if and only if~$\bx(\bn)\in\pd\cH$. Thus the
problem of minimizing~$\vep(\bn)$ is again equivalent to finding the distance (with respect to the
norm~$\Vert{\cdot}\Vert$) of the fixed vector~$\bh$ to the simplex~$\cH$. Since this simplex is a
convex polytope, there is a unique point in~$\cH$ closest to~$\bh$ for all values
of~$\bh\in\RR^m$. This proves that the magnon densities are uniquely determined by the magnetic
field strength~$\bh$, so that the ground state of the gLMG model~\eqref{HgLMG} is always unique.
More precisely, when~$\bh\in\Int\cH$, the distance of~$\bh$ to~$\cH$ is obviously zero and the
corresponding density vector~$\bn_0$ is again determined by the condition~$\bh=\bx(\bn_0)$, i.e.,
by Eq.~\eqref{nas}. On the other hand, when~$\bh$ lies outside~$\Int\cH$ (in particular,
if~$\bh\in\pd\cH$), it is clear from geometric considerations that the minimum distance of~$\bh$
to the simplex~$\cH$ is attained at a point~$\bx(\bn_0)\in\pd\cH$, so that the corresponding
magnon density~$\bn_0$ lies in~$\pd\cN$. Hence in the latter case the system is in one of the $m$
phases characterized by the vanishing of~$k>0$ magnon densities.

\section{Conclusions and outlook}\label{sec.conc}

In this work we have introduced a large class of $\su(m+1)$ spin models with long-range
interactions possessing a symmetric non-degenerate ground state with well-defined magnon numbers.
Our models provide a natural multiparameter generalization of the well-known spin~$1/2$ isotropic
Lipkin--Meshkov--Glick model, featuring non-constant interactions and~$\su(m+1)$ spin. They are
also closely related to Haldane--Shastry type chains, which can be formally obtained from them
through specific realizations of the couplings $h_{ij}$ dropping the magnetic field terms.

One of the main results of our work is the detailed derivation of the asymptotic behavior of the
eigenvalues of the reduced density matrix of a block of $L$ spins when the system is in its ground
state, for arbitrary values of $m$ and $\al=\lim_{N\to\infty}L/N$. This makes it possible to
compute the von Neumann and R\'enyi entropies in closed form when~$L\gg1$, and to derive their
asymptotic behavior when~$L\to\infty$. A notable outcome of our analysis is that both of these
entropies scale as~$\frac12(m-k)\log L$ in the latter limit, where $k$ is the number of vanishing
magnon densities in the ground state. In particular, from the behavior of the R\'enyi entropy it
follows that the class of generalized~$\su(m+1)$ LMG models contains no critical models. We have
also computed the Tsallis entropy, showing that it can be made extensive when the number of
``effective'' internal degrees of freedom~$m-k+1$ is greater than $3$. Finally, we have completely
determined the different phases of the ground state in terms of the~$\su(m+1)$ magnetic field
strength~$\bh$, and shown that they are related in a simple geometric way to the weights of the
fundamental representation of~$\su(m+1)$.

Our results open up several natural directions for further research and a number of related
problems. One such problem is the determination in closed form of the full spectrum of suitable
models of the class introduced in this paper, particularly for~$\su(m+1)$ spin with~$m>1$. In
fact, the integrability properties of the HS-type chains suggest the possibility of exploring the
existence of integrable generalizations thereof with a non-vanishing magnetic field term of the
form considered in this work. At the same time, it could also be of interest to extend the
analysis of the entanglement entropy performed in this paper to different entropic functionals.
Indeed, due to the presence of several parameters in the gLMG Hamiltonian, multiparametric
entropies~\cite{Te15,Te16} could play an important role in the classification of the possible
thermodynamic regimes admitted by the system when one varies the values of its parameters.
Finally, the fact that the von Neumann entanglement entropy of generalized LMG models is
proportional to~$\log L$ in certain regions of parameter space, though as we have seen does not
imply the existence of critical models, suggests that these regions may nevertheless contain
models with interesting non-generic properties worth investigating. This is certainly true in
the~$m=1$ case, since (for instance) the isotropic LMG model is gapless precisely in the
interval~$|h|<1$ for which the von Neumann entropy scales as~$\log L$~\cite{BJP82}.

\ack This work was supported in part by Spain's MINECO under grant no.~FIS2011-22566 and by the
Universidad Complutense de Madrid and Banco Santander under grant no.~GR3/14-910556. JAC would
also like to thank the Madrid township and the ``Residencia de Estudiantes'' for their financial
support. The authors would also like to thank the anonymous referees of a previous version
of this manuscript for their helpful remarks and suggestions.

\appendix
\section{Ground-state reduced density matrix for a block of spins}\label{app.rhoL}

In this appendix we shall compute from first principles the reduced density matrix~$\rho_L$ of a
block of $L$ spins of the~$\su(m+1)$ gLMG chain~\eqref{HgLMG} when the system is in its ground
state~\eqref{gLMGgs}, with magnon densities~$n_a=N_a/N$ determined (in the thermodynamic limit) by
Eq.~\eqref{nas}. This result, first obtained in Ref.~\cite{PSS05}, shall be used in
Section~\ref{sec.entropies} to evaluate several standard bipartite entanglement entropies in the
thermodynamic limit.

More precisely, we need to compute
\begin{equation}\label{rhoL}
\rho_L=\tr_{N-L}\rho\equiv\tr_{N-L}\ket{\psi(N_1,\dots,N_{m+1})}\bra{\psi(N_1,\dots,N_{m+1})}\,,
\end{equation}
where $\tr_{N-L}$ is the trace over the degrees of freedom of the remaining $N-L$ spins. Since the
ground state~$\ket{\psi(N_1,\dots,N_{m+1})}$ is invariant under permutations, the result is
obviously independent of the specific positions of the~$L$ spins considered. We shall therefore
assume in what follows that the two blocks under consideration consist of the first $L$ and the
last $N-L$ spins. In order to evaluate the RHS of Eq.~\eqref{rhoL}, it is convenient to label the
states of the canonical spin basis~$\ket{s_1,\dots,s_N}$ by the positions of its magnons. More
explicitly, we shall use the notation $\ket{\biv_1,\dots,\biv_m}$ to denote a state whose type~$a$
magnons (with $a=1,\dots,m$) are located in the positions specified by the $N_a$ components of the
ordered multi-index $\biv_a$, while those of type~$m+1$ lie in the remaining positions. Note, in
particular, that no two multi-indices $\biv_a$ and $\biv_b$ with $a\ne b$ can have any common
components, which we shall denote by~$\biv_a\cap\biv_b=\emptyset$. For instance, with this
notation the~$\su(4)$ basis state $ \ket{3,1,1,4,1,3,4}$ will be denoted by
$\ket{(2,3,5),(),(1,6)}$.

Using the above notation, the ground state~\eqref{gLMGgs} can be written as
\begin{equation}
  \label{gLMGis}
  \ket{\psi(N_1,\dots,N_{m+1})}=\left(\frac{N!}{\prod_{a=1}^{m+1}N_a!}\right)^{\!\!-\frac{1}{2}}
  \sum_{\biv_1,\dots,\biv_m}\ket{\biv_1,\dots,\biv_m}\,,
\end{equation}
where the sum is over all ordered multi-indices~$\biv_a\in\{1,\dots,N\}^{N_a}$ such
that~$\biv_a\cap\biv_b=\emptyset$ for $a\ne b$. We thus have
\[
\rho=\frac{\prod_{a=1}^{m+1}N_a!}{N!}\sum_{\substack{\biv_1,\dots,\biv_m\cr\bj_1,\dots,\,\bj_m}}
\ket{\biv_1,\dots,\biv_m}\bra{\,\bj_1,\dots,\bj_m}\,,
\]
where the sum is again over all ordered multi-indices~$\biv_a$, $\bj_a$ satisfying the above
condition. Thus, in order to evaluate~$\rho_L$ we need only
compute~$\tr_{N-L}\ket{\biv_1,\dots,\biv_m}\bra{\,\bj_1,\dots,\bj_m}$. To this end, we decompose
each multi-index~$\biv_a$ as
\[
\biv_a=(\biv_a^L,\biv_a^{N-L})\,,
\]
where the components of each ~$\biv_a^L$ range from~$1$ to $L$ and those of $\biv_a^{N-L}$ from
$L+1$ to $N$, and similarly for~$\bj_a$. It is then straightforward to show that
\begin{equation}\label{trNmLij}
  \fl
\tr_{N-L}\ket{\biv_1,\dots,\biv_m}\bra{\,\bj_1,\dots,\bj_m}
=\ket{\biv_1^L,\dots,\biv_m^L}\bra{\,\bj_1^L,\dots,\bj_m^L}\,\prod_{a=1}^m\de_{\biv_a^{N-L},\,\bj_a^{N-L}}\,.
\end{equation}
Indeed,
\[
\inner{s_{L+1},\dots,s_N}{\biv_1,\dots,\biv_m}\,\inner{\,\bj_1,\dots,\bj_m}{s_{L+1},\dots,s_N}=0
\]
unless the last $N-L$ spin components of the basis states represented
by~$\ket{\biv_1,\dots,\biv_m}$ and $\ket{\,\bj_1,\dots,\bj_m}$ are both equal to
$s_{L+1},\dots,s_N$, which accounts for the product of Kronecker deltas in Eq.~\eqref{trNmLij}.
Moreover, when~$\biv_a^{N-L}=\bj_a^{N-L}$ for all~$a=1,\dots,m$ the only state of the canonical
basis of the Hilbert space of the last $N-L$ spins that can contribute to the trace
is~$\ket{\biv_1^{N-L},\dots,\biv_m^{N-L}}=\ket{\bj_1^{N-L},\dots,\bj_m^{N-L}}$, which immediately
yields Eq.~\eqref{trNmLij}.

Using Eq.~\eqref{trNmLij} it is straightforward to obtain the expression
\begin{equation}\label{rhoLsum}
\rho_L=\frac{\prod_{a=1}^{m+1}N_a!}{N!}
\sum_{\substack{\biv_1^L,\dots,\biv_m^L,\,\bj_1^L,\dots,\,\bj_m^L\,,
    \biv_1^{N-L},\dots,\biv_m^{N-L}\cr|\biv_a^L|=|\bj_a^L|=N_a-|\biv_a^{N-L}|
  }}\ket{\biv_1^L,\dots,\biv_m^L}\bra{\,\bj_1^L,\dots,\bj_m^L}\,,
\end{equation}
where $|\biv|$ denotes the number of components of the multi-index $\biv$. In order to evaluate the
latter sum, we introduce the notation $L_a=|\biv_a^L|$, $a=1,\dots,m$, where the magnon numbers
$L_1,\dots,L_m$ satisfy the obvious inequalities
\begin{equation}\label{Laineqs}
0\le L_a\le N_a\,,\qquad \sum_{a=1}^mL_a\le L\,,\qquad \sum_{a=1}^m(N_a-L_a)\le N-L\,.
\end{equation}
Equation~\eqref{rhoLsum} can then be written as
\begin{equation}\label{rhoLsum2}
  \fl
\rho_L=\frac{\prod_{a=1}^{m+1}N_a!}{N!}
\sum_{L_1,\dots,L_m}\,\sum_{\substack{\biv_1^L,\dots,\biv_m^L,\,\bj_1^L,\dots,\,\bj_m^L\cr|\biv_a^L|
    =|\bj_a^L|=L_a
  }}\sum_{\substack{\biv_1^{N-L},\dots,\biv_m^{N-L}\cr|\biv_a^{N-L}|=N_a-L_a
  }}\ket{\biv_1^L,\dots,\biv_m^L}\bra{\,\bj_1^L,\dots,\bj_m^L}\,,
\end{equation}
where the outermost sum is over the range specified by Eq.~\eqref{Laineqs}. The sum over the
multi-indices~$\biv_a^{N-L}$ is clearly the number of different $(N-L)$-particles states of the
form~$\ket{\biv_1^{N-L},\dots,\biv_m^{N-L}}$ with a fixed number $N_a-L_a$ of type~$a$ magnons
($1\le a\le m+1$), namely the combinatorial number
\[
\frac{(N-L)!}{\prod_{a=1}^{m+1}(N_a-L_a)!}
\]
with $L_{m+1}\equiv L-\sum\limits_{a=1}^m L_a$. Thus Eq.~\eqref{rhoLsum2} reduces to
\[
\fl
\rho_L=\frac{\prod_{a=1}^{m+1}N_a!}{N!}
\sum_{L_1,\dots,L_m}\frac{(N-L)!}{\prod_{a=1}^{m+1}(N_a-L_a)!}
\sum_{\substack{\biv_1^L,\dots,\biv_m^L\cr|\biv_a^L|=L_a }}\ket{\biv_1^L,\dots,\biv_m^L}\cdot
\sum_{\substack{\bj_1^L,\dots,\,\bj_m^L\cr|\bj_a^L|=L_a }}\bra{\,\bj_1^L,\dots,\bj_m^L}\,.
\]
Using Eq.~\eqref{gLMGis} with $N$ and $N_a$ respectively replaced by $L$ and $L_a$ we finally
arrive at the following explicit formula for the reduced density matrix~$\rho_L$:
\begin{equation}
  \label{rhoLLs}
  \rho_L=\sum_{L_1,\dots,L_m}\la(L_1,\dots,L_m)\ket{\psi(L_1,\dots,L_{m+1})}\bra{\psi(L_1,\dots,L_{m+1})}\,,
\end{equation}
where the summation range is again given by~Eq.~\eqref{Laineqs} and
\begin{equation}
  \label{lambda-app}\fl
  \la(L_1,\dots,L_m)=\frac{L!}{\prod_{a=1}^{m+1}L_a!}\,\frac{(N-L)!}{\prod_{a=1}^{m+1}(N_a-L_a)!}\,
  \frac{\prod_{a=1}^{m+1}N_a!}{N!}=\binom NL^{\!\!-1}\prod_{a=1}^{m+1}\binom{N_a}{L_a}\,.
\end{equation}
Thus $\rho_L$ is diagonal in the basis~$\ket{\psi(L_1,\dots,L_{m+1})}$ with~$L_1,\dots,L_m$
satisfying~\eqref{Laineqs} (and $L_{m+1}=L-\sum_{a=1}^mL_a$), and its
eigenvalues~$\la(L_1,\dots,L_m)$ are given by Eq.~\eqref{lambda-app} (cf.~\cite{PSS05}). In the
particular case~$m=1$~\eqref{lambda-app} reduces to a hypergeometric distribution~\cite{Fe71}, as
shown in Ref.~\cite{LORV05} for the isotropic LMG model (see also~\cite{PS05}).

  According to the Schmidt decomposition theorem (see, e.g.,~\cite{NC10}), the ground
  state~$\ket{\psi(N_1,\dots,N_{m+1})}$ can be expressed as
  \begin{equation}\label{SD}
  \ket{\psi(N_1,\dots,N_{m+1})}=\sum_i b_i\ket{\psi_i}\otimes\ket{\vp_i}\,,
\end{equation}
where~$\{\ket{\psi_j}\}$ and $\{\ket{\vp_k}\}$ are appropriate orthonormal bases of the Hilbert
spaces of the first $L$ and last $N-L$ particles, and the Schmidt coefficients $b_i$ are
non-negative real numbers. From this formula it immediately follows that
  \[
  \rho_L=\sum_ib_i^2\,\ket{\psi_i}\bra{\psi_i}\,,
  \]
  and comparing with Eq.~\eqref{rhoLLs} we obtain
  \[
  \ket{\psi_i}=\ket{\psi(L_1,\dots,L_{m+1})}\,,\qquad b_i=\sqrt{\la(L_1,\dots,L_{m+1})}\,.
  \]
  In fact, in this case it is straightforward to derive the Schmidt decomposition~\eqref{SD}
  directly. Indeed, using the previous notation for the multi-indices~$\biv_a$ and
  Eq.~\eqref{gLMGis} we have
  \begin{eqnarray*}
    \fl
    &\ket{\psi(N_1,\dots,N_{m+1})}
    \\\fl
    &=\left(\frac{N!}{\prod_{a=1}^{m}N_a!}\right)^{\!\!-\frac{1}{2}}
      \sum_{\substack{\biv_1^L,\dots,\biv_m^L\cr\biv_1^{N-L},\dots,\,\biv_m^{N-L}}}\ket{\biv_1^L,\dots,\biv_m^L}\otimes
      \ket{\biv_1^{N-L},\dots,\biv_m^{N-L}}
    \\\fl
    &=\left(\frac{N!}{\prod_{a=1}^{m+1}N_a!}\right)^{\!\!-\frac{1}{2}}\sum_{L_1,\dots,L_m}
      \left(\sum_{\substack{\biv_1^L,\dots,\biv_m^L\cr|\biv_a^L|=L_a}}\ket{\biv_1^L,\dots,\biv_m^L}\right)
      \otimes
      \left(\sum_{\substack{\biv_1^{N-L},\dots,\biv_m^{N-L}\cr|\biv_a^{N-L}|=N_a-L_a}}
      \ket{\biv_1^{N-L},\dots,\biv_m^{N-L}}\right)
    \\\fl
    &=\sum_{L_1,\dots,L_m}\sqrt{\la(L_1,\dots,L_m)}\,\ket{\psi(L_1,\dots,L_{m+1})}\otimes
    \ket{\psi(N_1-L_1,\dots,N_{m+1}-L_{m+1})}\,.
  \end{eqnarray*}
  Again, in the~$\su(2)$ case the latter equation reduces to the analogous formula~in
  Ref.~\cite{LORV05}.

  It is also important to observe that all the results in this section are based exclusively on
  the fact that the ground state of the gLMG model~\eqref{HgLMG} is of the form~\eqref{gLMGgs},
  i.e., it is symmetric and has well-defined magnon densities. Thus the above results hold, in
  general, for \emph{any} quantum system whose ground state is of the latter form.
\section{Moments of the multivariate hypergeometric distribution}\label{app.moments}
In this appendix we shall compute in closed form the first and second moments of the multivariate
hypergeometric distribution~$\la(L_1,\dots,L_m)$ given by Eq.~\eqref{lambda}, with
support~\eqref{Laineqs}. Our starting point is the elementary identity
\[
\prod_{i=1}^{m+1}(1+t_i)^{N_i}=\sum_{L_1=0}^{N_1}\cdots\sum_{L_{m+1}=0}^{N_{m+1}}\prod_{i=1}^{m+1}\binom{N_i}{L_i}
\cdot t_1^{L_1}\cdots t_{m+1}^{L_{m+1}}\,,
\]
from which we deduce that
\begin{equation}\label{momentsid}
  \fl
  \sum_{L_1=0}^{N_1}\cdots\sum_{L_{m+1}=0}^{N_{m+1}}\prod_{k=1}^n(L_i-k+1)\cdot\prod_{j=1}^{m+1}\binom{N_j}{L_j}
  \cdot t_1^{L_1}\cdots
  t_{m+1}^{L_{m+1}}=t_i^n\pdf{^n}{t_i^n}\,\prod_{j=1}^{m+1}(1+t_j)^{N_j}.
\end{equation}
The previous formula can be applied to compute the moments of the
distribution~$\la(L_1,\dots,L_m)$ in a straightforward way. Indeed, let
\[
\langle f(L_1,\dots,L_m)\rangle\equiv\sum_{L_1,\dots,L_m}f(L_1,\dots,L_m)\la(L_1,\dots,L_m)\,,
\]
where the sum ranges over the set determined by the inequalities~\eqref{Laineqs}, denote the
average of the function~$f(L_1,\dots,L_m)$ with respect to the distribution~\eqref{lambda}.
Equation~\eqref{momentsid} with~$N_1+\cdots+N_{m+1}=N$ implies that
\begin{equation}
  \label{momsform}
  \fl
  \left\langle\prod_{k=1}^n(L_i-k+1)\right\rangle=
  \binom{N}{L}^{-1}\left\{t_i^n\pdf{^n}{t_i^n}\,
    \prod_{j=1}^{m+1}(1+t_j)^{N_j}\Big|_{t_1=\cdots
      =t_{m+1}=t}\right\}_L,
\end{equation}
where~$\{\phi(t)\}_L$ denotes the coefficient of $t^L$ in the polynomial~$\phi(t)$. From the
latter formula with~$n=1$ we obtain
\begin{equation}
  \label{Liav}
\langle L_i\rangle =
\binom{N}{L}^{-1}\left\{N_it(1+t)^{N-1}\right\}_L=N_i\,\frac{\binom{N-1}{L-1}}{\binom{N}{L}}
=\frac{LN_i}{N}=Ln_i\,.
\end{equation}
Similarly, Eq.~\eqref{momsform} with~$n=2$ yields
\begin{eqnarray*}
\big\langle L_i(L_i-1)\big\rangle &=
\binom{N}{L}^{-1}\left\{N_i(N_i-1)t^2(1+t)^{N-2}\right\}_L=N_i(N_i-1)\,\frac{\binom{N-2}{L-2}}{\binom{N}{L}}\\
  &=\frac{L(L-1)N_i(N_i-1)}{N(N-1)}\,,
\end{eqnarray*}
so that
\begin{equation}
  \fl
  \big\langle x_i^2\big\rangle\equiv
\Big\langle(L_i-\langle L_i\rangle)^2\Big\rangle
=\big\langle L_i(L_i-1)\big\rangle-\langle L_i\rangle\big(\langle L_i\rangle-1\big)
=Ln_i(1-n_i)\,\frac{N-L}{N-1}\,.
  \label{xi2av}
\end{equation}
In particular, in the thermodynamic limit~$N\to\infty$ with~$\lim_{N\to\infty}(L/N)=\al$ finite we
obtain the asymptotic formula
\begin{equation}
  \label{xi2avasymp}
  \big\langle x_i^2\big\rangle\simeq L(1-\al)n_i(1-n_i)\,.
\end{equation}
The latter equation generalizes the analogous formula derived in Ref.~\cite{PSS05} by
approximating $\la(L_1,\dots,L_m)$ by a multinomial distribution, valid only for~$\al=0$. Finally,
the covariances~$\langle x_ix_j\rangle=\langle L_iL_j\rangle-\langle L_i\rangle\langle L_j\rangle$
with $i\ne j$ can also be easily evaluated from the identity
\begin{equation}
  \label{covsform}
  \left\langle L_iL_j\right\rangle=
  \binom{N}{L}^{-1}\left\{t_it_j\pdf{}{t_i}\pdf{}{t_j}\,
    \prod_{k=1}^{m+1}(1+t_k)^{N_k}\Big|_{t_1=\cdots
      =t_{m+1}=t}\right\}_L,
\end{equation}
whose proof is similar to that of Eq.~\eqref{momsform}. From Eq.~\eqref{covsform} we easily obtain
\[
\fl
\left\langle L_iL_j\right\rangle=\binom{N}{L}^{-1}\left\{N_iN_jt^2(1+t)^{N-2}\right\}_L
=N_iN_j\frac{\binom{N-2}{L-2}}{\binom{N}{L}}\\
=\frac{L(L-1)N_iN_j}{N(N-1)},\qquad i\ne j\,,
\]
and hence, by Eq.~\eqref{Liav},
\begin{equation}
  \label{covs}
  \big\langle x_ix_j\big\rangle=\frac{L(L-1)N_iN_j}{N(N-1)}-\frac{L^2N_iN_j}{N^2}
  =-Ln_in_j\,\frac{N-L}{N-1}\,,\qquad i\ne j\,.
\end{equation}
Again, in the thermodynamic limit we obtain the asymptotic formula
\begin{equation}
  \label{xixjavasymp}
  \big\langle x_ix_j\big\rangle\simeq -L(1-\al)n_in_j\,,\qquad i\ne j\,,
\end{equation}
which for~$\al=0$ yields the analogous formula in Ref.~\cite{PSS05}.
\section{Limit of the hypergeometric probability distribution}
\label{app.hypgeom} 

In this appendix we shall provide a brief self-contained proof of the approximation of a
hypergeometric probability distribution by a suitable Gaussian distribution, used in
Section~\ref{sec.entropies} to derive the behavior of the eigenvalues of the reduced density
matrix~$\rho_L$ in the thermodynamic limit.

Consider the hypergeometric probability distribution
\begin{equation}\label{hypdis}
p_l=\frac{\binom{\tilde L}l\binom{\tilde N-\tilde L}{n-l}}{\binom{\tilde N}n}\,,\qquad
l=0,1,\dots, \tilde L\,,
\end{equation}
where $0\le n,\tilde L\le \tilde N$ are fixed. We are interested in approximating $p_l$
when~$\tilde N\to\infty$, assuming that
$\lim_{\tilde N\to\infty}(\tilde L/\tilde N)\equiv\tilde\al$ and
$\lim_{\tilde N\to\infty}(n/\tilde N)\equiv\nu$ with~$0<\tilde\al,\nu<1$. To this end, note that
we can write
\begin{equation}\label{plx}
  p_l=\frac{\binom{\tilde L}lx^l(1-x)^{\tilde L-l}\cdot
    {\binom{\tilde N-\tilde L}{n-l}x^{n-l}
      (1-x)^{\tilde N-\tilde L-n+l}}}{{\binom{\tilde N}nx^n(1-x)^{\tilde N-n}}}
\end{equation}
for arbitrary~$x\in(0,1)$. According to the Laplace--de Moivre theorem, for~$K\gg1$ a binomial
distribution
\[
\binom Kk x^k(1-x)^{K-k}
\]
can be approximated by the continuous Gaussian distribution
\begin{equation}\label{Gauss}
g(k;\mu,\si)\equiv\frac{\e^{-\frac{(k-\mu)^2}{2\si^2}}}{\sqrt{2\pi\si^2}}\,,
\end{equation}
where
\[
\mu=xK\,,\qquad\si^2=Kx(1-x)\,.
\]
By Eqs.~\eqref{Liav} and~\eqref{xi2avasymp} with $m=1$, in the limit considered the mean and
variance of the random variable~$l$ are given by
\begin{equation}\label{lamuhyp}
  \langle l\rangle=\nu\tilde L\,,\qquad \langle l^2\rangle-\langle l\rangle^2\simeq\tilde L(1-\tilde\al)\nu(1-\nu)\,.
\end{equation}
Hence
$\big(\langle l^2\rangle-\langle l\rangle^2\big)^{1/2}/\langle l\rangle=\Or(\tilde L^{-1/2})$, so
that the hypergeometric distribution~\eqref{hypdis} is sharply peaked around its
average~$\nu\tilde L$. It is immediate to check that when $l$ is near~$\nu \tilde L$ we can
simultaneously approximate the three binomial distributions in Eq.~\eqref{plx} using the
Laplace--de Moivre formula with $x=\nu$ and suitable choices of~$k$ and $K$. We thus obtain
\begin{equation}\label{plg}
  \fl p_l\simeq\sqrt{2\pi \tilde N\nu(1-\nu)}\cdot \frac{\e^{-\frac{(l-\tilde N\tilde\al
        \nu)^2}{2\tilde N\tilde\al\nu(1-\nu)}}}{\sqrt{2\pi \tilde N\tilde\al\nu(1-\nu)}}\,\cdot
        \frac{\e^{-\frac{(\tilde N\nu-l-\tilde N(1-\tilde\al)\nu)^2}{2\tilde N(1-\tilde\al)\nu(1-\nu)}}}{\sqrt{2\pi
    \tilde N(1-\tilde\al)\nu(1-\nu)}}=g(l;\mu,\si)\,,
\end{equation}
with
\begin{equation}
  \label{musigma}
  \mu=\tilde N\tilde\al\nu\,,\qquad \si^2=\tilde N\tilde\al(1-\tilde\al)\nu(1-\nu)\,.
\end{equation}
Note, finally, that these values of~$\mu$ and~$\si^2$ respectively coincide with the asymptotic
values of the mean and variance of the random variable~$l$ (cf.~Eq.~\eqref{lamuhyp}).

\section*{References}


\end{document}